%% file: main.tex
\RequirePackage{fix-cm}
\documentclass[smallextended,numbook]{svjour3}       %
\smartqed  %
\usepackage{graphicx}

\input{config/preamble}

\input{config/commands}
\input{config/acronyms}

\journalname{Empirical Software Engineering}

\begin{document}

\title{Breaking Bad? Semantic Versioning and Impact of Breaking Changes in Maven Central}
\subtitle{An external and differentiated replication study}

\titlerunning{Semantic Versioning and Impact of Breaking Changes in Maven Central}        %

\author{Lina Ochoa \and
	Thomas Degueule \and
	Jean-Rémy Falleri \and
	Jurgen Vinju
}

\institute{Lina Ochoa \at
	\email{l.m.ochoa.venegas@tue.nl} \\
	Eindhoven University of Technology, Eindhoven, The Netherlands \\ 
	\and
	Thomas Degueule \at
	\email{thomas.degueule@labri.fr} \\
	Univ. Bordeaux, Bordeaux INP, CNRS, LaBRI, Bordeaux, France \\
	\and
	Jean-Rémy Falleri \at
	\email{falleri@labri.fr} \\
	Univ. Bordeaux, Bordeaux INP, CNRS, LaBRI, Bordeaux, France \\
	Institut Universtaire de France \\
	\and 
	Jurgen Vinju \at
	\email{jurgen.vinju@cwi.nl} \\
	Centrum Wiskunde \& Informatica, Amsterdam, The Netherlands \\
	Eindhoven University of Technology, Eindhoven, The Netherlands \\ 
}

\date{Received: date / Accepted: date}

\maketitle

\begin{abstract}
Just like any software, libraries evolve to incorporate new features, bug fixes, security patches, and refactorings.
However, when a library evolves, it may break the contract previously established with its clients by introducing Breaking Changes (BCs) in its API.
These changes might trigger compile-time, link-time, or run-time errors in client code.
As a result, clients may hesitate to upgrade their dependencies, raising security concerns and making future upgrades even more difficult.

Understanding how libraries evolve helps client developers to know which changes to expect and where to expect them, and library developers to understand how they might impact their clients.
In the most extensive study to date, Raemaekers~\etal investigate to what extent developers of Java libraries hosted on the Maven Central Repository (MCR) follow semantic versioning conventions to signal the introduction of BCs and how these changes impact client projects.
Their results suggest that BCs are widespread without regard for semantic versioning, with a significant impact on clients.

In this paper, we conduct an external and differentiated replication study of their work.
We identify and address some limitations of the original protocol and expand the analysis to a new corpus spanning seven more years of the MCR.
We also present a novel static analysis tool for Java bytecode, \mrc, which provides us with:~(i)~the set of all BCs between two versions of a library; and (ii)~the set of locations in client code impacted by individual BCs.

Our key findings, derived from the analysis of $119,879$ library upgrades and $293,817$ clients, contrast with the original study and show that 83.4\% of these upgrades do comply with semantic versioning.
Furthermore, we observe that the tendency to comply with semantic versioning has significantly increased over time.
Finally, we find that most BCs affect code that is not used by any client, and that only 7.9\% of all clients are affected by BCs.
These findings should help (i) library developers to understand and anticipate the impact of their changes; (ii) library users to estimate library upgrading effort and to pick libraries that are less likely to break; and (iii) researchers to better understand the dynamics of library-client co-evolution in Java.
\end{abstract}

\input{content/intro}
\input{content/background}

\input{content/original-study}
\input{content/protocol}
\input{content/analysis}
\input{content/rw}

\input{content/discussion}

\input{content/conclusion}

\begin{acknowledgements}
This research has been carried out as part of the CROSSMINER project, which has received funding from the European Union’s Horizon 2020 Research and Innovation Programme under grant agreement No.~732223. The authors would like to thank Steven Raemaekers for sharing all the scripts and data of the replicated work, and the anonymous reviewers for their perceptive comments.
\end{acknowledgements}

\section*{Declarations}

\paragraph{Funding}
This work was partially supported by the EU’s Horizon 2020 Project No. 732223 CROSSMINER.

\paragraph{Conflict of interest}
None.

\paragraph{Data and code availability}
In order to ease the replication of our results, we make available the datasets, and the code used to generate and analyse them in a single replication package hosted on Zenodo:

\begin{center}
	\url{https://zenodo.org/record/5221840}
\end{center}

The archive contains the software project used to generate the datasets together with the instructions to run the code. 
The study analysis is also included in the project as a set of Jupyter notebooks.
It also gives information on how to access the MDD and MDG corpora.
Relevant links and sources are specified within the README files and the companion webpage.

\bibliographystyle{spbasic}      %
\bibliography{main}   %

\end{document}

%% file: config/preamble.tex
\usepackage[T1]{fontenc} %
\usepackage[utf8]{inputenc} %
\usepackage[all]{nowidow}
\usepackage{xcolor}
\usepackage{listings}
\usepackage{balance} %
\usepackage{microtype} %
\usepackage[colorlinks=true,allcolors=black,breaklinks]{hyperref}  %
\usepackage[numbers]{natbib} %
\usepackage[inline]{enumitem} %
\usepackage{tikz}
\usepackage[scaled=.7]{beramono}
\usepackage{booktabs} %
\usepackage{multirow} %
\usepackage{rotating} %
\usepackage{filecontents}
\usepackage{siunitx}
\usepackage{graphicx}
\usepackage{xspace} %
\usepackage{mdframed}
\usepackage{amssymb}
\usepackage{url}
\usepackage{amsmath}
\usepackage{subfloat}
\usepackage{caption}
\usepackage[abbreviations]{foreign}
\usepackage{subcaption}
\usepackage{pdflscape}  
\usepackage{cleveref} %

\usetikzlibrary{arrows, arrows.meta, chains, positioning, shapes, patterns}

\tikzstyle{base} = [draw, thick, align=center, inner ysep=1mm, inner xsep=2mm]
\tikzstyle{startstop} = [draw, rounded rectangle, text centered, draw=black,thick]
\tikzstyle{io} = [trapezium, base, trapezium left angle=70, trapezium right angle=110, text centered]
\tikzstyle{process} = [rectangle, text centered, draw=black,thick]
\tikzstyle{decision} = [diamond, text centered, draw=black,thick]
\tikzstyle{arrow} = [-{Stealth[scale=1.2]},rounded corners,thick]

\newenvironment{Result}[1][]{%
	\begin{mdframed}[%
		frametitle={#1},
		skipabove=\baselineskip plus 2pt minus 1pt,
		skipbelow=\baselineskip plus 2pt minus 1pt,
		linewidth=0.5pt,
		frametitlerule=true,
		frametitlebackgroundcolor=gray!30,
		nobreak=true
		]%
	}{%
	\end{mdframed}
}

%% file: config/commands.tex
\definecolor{keywordscolor}{RGB}{127, 0, 85}
\definecolor{stringcolor}{RGB}{42, 0, 255}
\definecolor{commentscolor}{RGB}{63, 127, 95}
\definecolor{annotationscolor}{RGB}{100, 100, 100}
\definecolor{lstbgcolor}{RGB}{245, 245, 245}

\newcommand*{\pom}[1]{\lstinline[language=pom]{#1}} %
\lstdefinelanguage{pom}{
	basicstyle=\ttfamily,
	stringstyle=\color{black},
	morekeywords={project, modelVersion, groupId, artifactId, version, dependencies, dependency, packaging, scope}
}

\newcommand*{\ijava}[1]{\lstinline[basicstyle=\ttfamily,language=Java,morekeywords={record,strictfp}]{#1}}

%% file: config/acronyms.tex
\usepackage{acronym} %

\acrodef{ADT}[ADT]{Algebraic Data Type}
\acrodef{API}[API]{Application Programming Interface}
\acrodef{BC}[BC]{Breaking Change}
\acrodef{EDA}[EDA]{Exploratory Data Analysis}
\acrodef{IoC}[IoC]{Inversion of Control}
\acrodef{JLS}[JLS]{Java Language Specification}
\acrodef{MCR}[MCR]{Maven Central Repository}
\acrodef{MDD}[MDD]{Maven Dependency Dataset}
\acrodef{MDG}[MDG]{Maven Dependency Graph}
\acrodef{POM}[POM]{Project Object Model}
\acrodef{SIG}[SIG]{Software Improvement Group}

\newcommand*{\mrc}{\texttt{Maracas}\@\xspace}
\newcommand*{\japi}{\texttt{japicmp}\@\xspace}
\newcommand*{\clirr}{\texttt{clirr}\@\xspace}
\newcommand*{\semver}{\texttt{semver}\@\xspace}
\newcommand*{\mt}{$M^3$\@\xspace}
\newcommand*{\rqone}{\textbf{Q1}\@\xspace}
\newcommand*{\rqtwo}{\textbf{Q2}\@\xspace}
\newcommand*{\rqthree}{\textbf{Q3}\@\xspace}
\newcommand*{\hone}{\textbf{H\textsubscript{1}}\@\xspace}
\newcommand*{\htwo}{\textbf{H\textsubscript{2}}\@\xspace}

\newcommand*{\dmodel}{$\Delta$-model\@\xspace}
\newcommand*{\dmodels}{$\Delta$-models\@\xspace}

\newcommand*{\duo}{$\mathcal{D}_u^o$\@\xspace}
\newcommand*{\dur}{$\mathcal{D}_u^r$\@\xspace}

\newcommand*{\dddo}{$\mathcal{D}_d^o$\@\xspace}
\newcommand*{\dddr}{$\mathcal{D}_d^r$\@\xspace}

\newcommand*{\os}{original study\@\xspace}

%% file: content/intro.tex
\section{Introduction}
\label{sec:intro}

Just like any software, libraries evolve to incorporate new features, bug fixes, security patches, and refactorings.
It is critical for clients to stay up to date with the libraries they use to benefit from these improvements and to avoid technical lag~\cite{gonzalez2017technical, ZeroualiMGDCR19}.
When a library evolves, however, it may break the contract previously established with its clients by introducing \acp{BC} in its public \ac{API}, resulting in compilation-time, link-time, or run-time errors.
These errors burden client developers given the sudden urgency to fix issues without intrinsic motivation.
As a result, clients may hesitate to upgrade their dependencies, raising security concerns and making future upgrades even more difficult~\cite{gaikovina2017developers,mirhosseini2017automated}.

\acp{BC} are language-specific:~they vary with the syntax and semantics of a particular programming language.
In Java, seemingly innocuous changes such as altering the visibility or abstractness modifier of a type declaration, or simply inserting a new method into an abstract type can, under certain conditions, break client code~\cite{Gosling2014Java}.
Most refactoring operations, though essential to maintain and evolve libraries, also induce \acp{BC}.
Thus, it does not come as a surprise that \acp{BC} are widespread in Java libraries~\cite{Xavier2017Historical}.
It is, however, essential to realize that not all \acp{BC} are intrinsically harmful.
Nonetheless, they should not come unannounced and take clients by surprise:~it should be clear for clients what consequences upgrading their dependencies will have on their own software, so they can make an informed decision beforehand.

To this end, Java library developers can leverage various mechanisms to communicate with their clients on the stability of their APIs and the effort required to upgrade to a newer version.
These mechanisms enable them to specify \emph{when} and \emph{where} \acp{BC} are to be expected.
On the one hand, semantic versioning (\semver) enables developers to use well-defined versioning conventions to classify new library releases as \emph{major} releases (which may introduce \acp{BC}), \emph{minor} releases (which may introduce new backward-compatible features but should not introduce any \ac{BC}), \emph{patch} releases (which should not affect the public API whatsoever), and \emph{initial development} releases (which may break anything at any time)~\cite{Preston2013Semver}.
On the other hand, annotations directly placed on source code elements (\eg Google's \texttt{@Beta} and Apache's \texttt{@Internal}) and naming conventions (such as \textit{internal} and \textit{experimental} packages) can be used to indicate that certain parts of the public API are exempt from compatibility guarantees and subject to sudden changes.

Clients who upgrade towards a new major release of a library or early adopters who rely on beta-stage APIs are well aware of the consequences.
It is thus crucial to distinguish between libraries that evolve gracefully by introducing \acp{BC} only when and where appropriate, and those that ``break bad'' by introducing \acp{BC} in minor and patch releases or in allegedly stable APIs.

In the most extensive study to date, Raemaekers et al. dissect backwards compatibility issues in the Maven Central Repository\footnote{\url{https://search.maven.org}} (MCR) with respect to semantic versioning~\cite{Raemaekers2017Semantic}.
The study uses the tool \clirr to infer the list of BCs between two versions of a Java library and measures their impact on client code using the Java compiler itself.
The empirical evaluation is carried on a complete snapshot of \acs{MCR}, up to the year 2011.
To name but a few of their findings, the study concludes that: 
\begin{enumerate*}[label=(\roman*)]
	\item \acp{BC} are widespread without regard for versioning conventions;
	\item the adherence to semantic versioning principles has increased over time, and;
	\item \acp{BC} have a significant impact on clients. 
\end{enumerate*}
The relevance and quality of this study for understanding the API-client co-evolution problem motivate us to replicate and expand its protocol and corpus.

In this paper, we conduct an external and differentiated replication study~\cite{Lindsay1993Design} of the study by \citet{Raemaekers2017Semantic}, which from now on we will refer to as the \emph{original study}.
After reviewing the original protocol, we introduce major changes to alleviate some of its limitations and address key threats to its validity. %
The main differences between our study and the \os are as follows:
\begin{itemize}[leftmargin=*]
	\item We refine the original protocol by introducing new filters and sanity checks to avoid analysing Maven artefacts that are not used as libraries and versions that are not meant to be used by clients---only $12\%$ of all artefacts in our replication corpus are indeed used as libraries;
	\item We implement a new tool built atop \japi,\footnote{\url{https://siom79.github.io/japicmp}} \mrc, more accurate than \clirr, which we use to analyse Java bytecode and compute the set of BCs between two versions of a library, as well as to compute how client projects are impacted by individual changes;
	\item We re-analyse the original corpus to assess the impact of our new protocol and tool, and expand the analysis to a new corpus spanning seven more years of the MCR (from $144$K Maven artefacts to $2.4$M).
\end{itemize}

We focus on a subset of three of the research questions investigated in the original study which are central to the API-client co-evolution problem, eluding other less relevant questions related to deprecation tags and characteristics of libraries that break more, among others.
Our research questions are as follows:
\begin{description}
	\item[\rqone] How are semantic versioning principles applied in the Maven Central repository in terms of \acp{BC}?
	\item[\rqtwo] To what extent has the adherence to semantic versioning principles increased over time?
	\item[\rqthree] What is the impact of \acp{BC} on clients? 
\end{description}

Our results show that, overall, library and client projects on \acl{MCR} are \emph{not} ``breaking bad''. 
First, 83.4\% of all library upgrades comply with \semver principles, introducing BCs only when they are expected. 
However, 20.1\% of non-major releases are breaking, being a potential threat to their clients.
Second, the tendency to comply with \semver practices has significantly increased over time.
In particular, the ratio of non-major releases introducing \acp{BC} has gradually decreased from 67.7\% in 2005 to 16.0\% in 2018.
Third, only 7.9\% of clients are actually impacted by BCs introduced in library releases. 
In most cases, clients do not use the breaking declarations (\ie the library declarations affected by \acp{BC})---but when they do, they are very likely to break. 
These results should help library developers to understand and anticipate the impact of their changes; library users to estimate library upgrading effort and to pick libraries that are less likely to break, and; researchers to better understand the dynamics of client-library co-evolution in Java and prioritize research in the future.

The remainder of this paper is organized as follows.
We first introduce background notions on Maven, \semver, and backwards compatibility in Java in \Cref{sec:background}.
We then briefly present the original study in \Cref{sec:orig}.
In \Cref{sec:protocol}, we detail the key differences in the protocol and datasets for our replication study.
We discuss our new results in \Cref{sec:analysis} and then present related work in \Cref{sec:rw}.
We then discuss the key findings and implications of our study in \Cref{sec:discussion} and finally conclude the paper and discuss future work in \Cref{sec:conclu}.

%% file: content/background.tex
\section{Background}
\label{sec:background}
In this section, we first introduce some background notions on Apache Maven, \acp{API}, and backwards compatibility in Java.
We also discuss the mechanisms available to developers to communicate the stability of their libraries through versioning conventions and source code annotations.

\subsection{Apache Maven}
Apache Maven (simply referred to as Maven hereafter) is a build automation tool particularly popular in the Java ecosystem.
Maven follows a plugin-oriented architecture that enables developers to specify the dependencies of a particular piece of software and how to build it.
When used to build Java projects, it enables developers to convert Java source code to Java bytecode (\texttt{.class} files) typically bundled as Java ARchives (JARs), which potentially depend on other JARs.
These artefacts can be deployed to and retrieved from remote Maven repositories.
The most popular Maven repository is the \ac{MCR} which, as of May 2021, hosts $6,723,367$ artefacts.

The cornerstone file defining a Maven project is the \ac{POM} file.
Typically, the \ac{POM} file is an XML file that contains metadata about the current project, its dependencies, and additional configurations required to build it.
\Cref{lst:pom} illustrates the typical structure and tags defined within a POM file, using the Spring TestContext Framework as an example.
The \pom{modelVersion} tag specifies the POM version of the file;~the \pom{groupId} tag identifies the organization or group that develops the project (\pom{org.springframework});~the  \pom{artifactId} tag identifies the project itself (\pom{spring-test});~the \pom{version} tag specifies the current version of the project (\pom{4.2.5.RELEASE});~and the \pom{packaging} tag specifies how the project is packaged (\pom{jar}).
Together, the group, artefact, and version (also known as \emph{project coordinates} and denoted \pom{groupId:artifactId:version}) uniquely identify a Maven artefact.

Dependencies of a project are declared within the \pom{dependencies} tag.
Each \pom{dependency} points to a unique Maven artefact (using its project coordinates), possibly supplemented with additional metadata.
In particular, the \pom{scope} tag specifies when the dependency is needed and thus in which classpath(s) it is included (\eg compile-time, test-time, or run-time dependencies).
One can automatically determine which libraries a Maven artefact depends on by parsing its POM file.
In \Cref{lst:pom}, the Spring TestContext Framework declares a compile-time dependency towards the JavaServlet library version \pom{3.0.1}.
Note that dependencies may employ version constraints (\eg \texttt{[1.0, 2.0)}), letting the dependency resolver find a suitable version within this range.

\begin{lstlisting}[language=pom, caption={Excerpt of the POM file of the Spring TestContext Framework project version \pom{4.2.5.RELEASE}.}, label={lst:pom},]
<project>
   <modelVersion>4.0.0</modelVersion>
   <groupId>org.springframework</groupId>
   <artifactId>spring-test</artifactId>
   <version>4.2.5.RELEASE</version>
   <packaging>jar</packaging>
   <dependencies>
      <dependency>
         <groupId>javax.servlet</groupId>
         <artifactId>javax.servlet-api</artifactId>
         <version>3.0.1</version>
         <scope>compile</scope>
      </dependency>
   </dependencies>
</project>
\end{lstlisting}

\subsection{API Evolution \& Backwards Compatibility}
\label{sec:api-evol}
An \acf{API} is an interface that exposes the set of services from a library that can be invoked by client projects.
In Java and other object-oriented languages, this interface consists of programming constructs such as \textit{packages}, \textit{types}, \textit{methods}, and \textit{fields}.
To delimit this interface, library developers use visibility modifiers and other dedicated constructs provided by the host language~\cite{Dig2006How}.

As an environment changes, software used in such an environment face the need to change accordingly.
This is what \citet{Lehman1978Programs} coined as \textit{software evolution}, later formalized as the eight Lehman's laws that synthesizes observations about software evolution~\cite{Godfrey2014Evolution, Lehman1997Metrics}.
Consequently, APIs---being software themselves---undergo continual and progressive change over time. 
The motivation behind this evolution is to provide more value to users by patching security issues, adding new features, simplifying the current API, fixing bugs, and improving maintainability~\cite{Dietrich2014Broken, Kula2018Empirical}. %

\begin{figure}[bt]
	\begin{subfigure}{.5\textwidth}
		\begin{lstlisting}[language=java]
public interface HttpServletRequest
  extends ServletRequest {
    public String getAuthType();
    public String getMethod();
	
    [...]
}
		\end{lstlisting}
		\caption{JavaServlet version \pom{3.0.1}.}
	\end{subfigure}
	\hfill
		\begin{subfigure}{.5\textwidth}
		\begin{lstlisting}[language=java]
public interface HttpServletRequest
  extends ServletRequest {
    public String getAuthType();
    public String getMethod();
    public String changeSessionId();
    [...]
}
		\end{lstlisting}
		\caption{JavaServlet version \pom{3.1.0}.}
	\end{subfigure}
	\hfill
		\begin{subfigure}{1.0\textwidth}
		\begin{lstlisting}[language=java]
public class MockHttpServletRequest implements HttpServletRequest {
	@Override public String getAuthType() {
		return this.authType;
	}
	@Override public String getMethod() {
		return this.method;
	}
	// MockHttpServletRequest must implement method HttpServletRequest.changeSessionId()
}
		\end{lstlisting}
		\caption{Spring TestContext Framework version \pom{4.2.5.RELEASE}.}
	\end{subfigure}
	\caption{Breaking change example:~Adding a new abstract method in the \ijava{HttpServletRequest} interface will break the client type \ijava{MockHttpServletRequest}.}
	\label{lst:bc-example}
\end{figure}

API evolution comes with the introduction of changes that can be classified according to how they affect client projects~\cite{Dig2006How} and specifically whether they ensure \textit{backwards compatibility}.
In Java, backwards compatibility is defined at the source, binary, and behavioural levels~\cite{Dietrich2014Broken}.
\textit{Source compatibility} is checked by the compiler when recompiling a client project with the new version of an API.
\textit{Binary compatibility} is checked by the Java Virtual Machine (JVM) during the linking process, as described in Chapter~13 of the \ac{JLS}~\cite{DesRivieres2007Evolving, Gosling2014Java}.
Lastly, \textit{behavioural compatibility} can only be verified at run time to check whether the program exhibits a behaviour that is different from its previous version, without triggering compilation or linkage errors~\cite{Dietrich2014Broken}.

There are two types of API changes, namely breaking and non-breaking changes.
On the one hand, \emph{breaking changes} (BCs) are not backwards compatible:~client projects using an API entity affected by a BC might break when migrating to a more recent version of the API~\cite{Dig2006How}.
On the other hand, \emph{non-breaking changes} (NBC) are backwards compatible, meaning that they do not trigger any source, binary, or behavioural incompatibility.
If an API only introduces backwards compatible changes, it is said to be \textit{stable}.
It is important to note that some BCs break several kinds of compatibility (\eg removing a public method is both source and binary incompatible), but none is a superset of the other~\cite{Jezek2015How}.
In this paper, to align with the original study, we only consider \emph{binary compatibility} and the associated set of BCs.

To illustrate how backwards incompatible changes might impact client projects, we refer to the Spring TestContext Framework example.
The JavaServlet library releases version \pom{3.1.0} in April 2013. 
This happens almost two years after its latest major release (\ie \pom{3.0.1}) in July 2011.
This new version introduces backwards incompatible changes that might break client code.
Some of those changes include adding new abstract methods to classes and interfaces.
These changes can potentially impact the Spring TestContext Framework in its \pom{4.2.5.RELEASE} version.
In some cases, stating that a BC affects client code is straightforward.
For instance, removing a type, method, or field that is used by a client will obviously break this client.
However, client code may also break when inserting new declarations in the library, for instance when inserting a new abstract method in an interface.
This change will break client code if it extends this interface, as illustrated in \Cref{lst:bc-example}.
In this case, the \ijava{changeSessionId()} method is added to the \ijava{HttpServletRequest} class within the JavaServlet library.
Given that the \ijava{MockHttpServletRequest} class in the Spring TestContext Framework implements such an interface, it will be forced to implement the new abstract method resulting in broken code.
The literature often overlooks the BCs induced by uses of a library in an \ac{IoC} style (\ie where the client extends types exposed in the library, following the Hollywood principle ``\emph{don't call us, we'll call you!}'')~\cite{Brito2018Why, Xavier2017Historical}.
In contrast, we include all of those in our analyses.
An exhaustive list of the 31 BCs we consider in this paper is available on the companion webpage.\footnote{\url{https://crossminer.github.io/maracas/detections}}

\subsection{API Stability Conventions}

It is critical for clients to be able to pinpoint which versions and which parts of an API introduce changes that might break their code.
\textit{Semantic versioning}, also known as \semver, is a popular convention to announce the introduction of \acp{BC}, and its use is encouraged in many software ecosystems (\eg npm, RubyGems, Cargo, Maven Central)~\cite{decan2019what}.
This versioning scheme is used to label library versions according to compatibility guarantees.
Each version number is specified in the form \texttt{<major>.<minor>.<patch>}, where \texttt{major}, \texttt{minor}, and \texttt{patch} are non-negative integers.
A change in the \texttt{major} version signals the possible introduction of backwards-incompatible changes.
Changes in the \texttt{minor} or \texttt{patch} versions signal the introduction of new features or bug fixes in a backwards-compatible fashion~\cite{Preston2013Semver}.
Initial development releases, which use zero as the \texttt{major} version, should also be considered unstable (``\emph{[m]ajor version zero (\texttt{0.Y.Z}) is for initial development. Anything MAY change at any time. The public API SHOULD NOT be considered stable}''~\cite{Preston2013Semver}).
Finally, version numbers may be suffixed with hyphen-separated qualifiers specifying pre-releases or build metadata (\eg \texttt{2.1.1-beta2}).

At the code level, library developers may use annotations such as Google's \texttt{@Beta} and Apache's \texttt{@Internal} to signal unstable declarations.
For instance, Guava developers state that ``\textit{APIs marked with the \texttt{@Beta} annotation at the class or method level are subject to change. They can be modified in any way, or even removed, at any time},''\footnote{\url{https://guava.dev/\#important-warnings}} and Apache POI developers state that ``\textit{Program elements annotated \texttt{@Internal} are intended for [...] internal use only. Such elements are not public by design and likely to be removed, have their signature change, or have their access level decreased [...] without notice}.''\footnote{\url{https://poi.apache.org/apidocs/dev/org/apache/poi/util/Internal}}
Naming conventions on packages (\eg \textit{internal} and \textit{experimental} packages) are sometimes used for the same purpose~\cite{businge2019stable}.
For instance, the following comment is attached to the class \texttt{Finalizer} contained in the package \texttt{com.google.common.base.internal} of Guava: ``\textit{While this class is public, we consider it to be *internal* and not part of our published API. It is public so we can access it reflectively across class loaders in secure environments}''.\footnote{\url{https://guava.dev/releases/9.0/api/docs/com/google/common/base/internal/Finalizer}}
This comment highlights the lack of mechanisms for developers to fine-tune the boundaries of their APIs in languages such as Java.
Some elements are made public because of technical constraints and not because of the desire to expose these elements in the API;~developers must therefore rely on band-aid solutions such as naming conventions.
When used in relation to \semver, these code-level mechanisms enable library developers to delimit a portion of their API that escapes the strict rules regarding backwards compatibility.
That is, BCs can be introduced in declarations labelled with these mechanisms without regard for \semver.

%% file: content/original-study.tex
\section{Original Study}
\label{sec:orig}
In this section, we briefly introduce the original study object of this replication.
We present its goal, main findings, and the protocol used to answer its research questions.

The original study by Raemaekers, van Deursen, and Visser, entitled ``\textit{Semantic versioning and impact of breaking changes in the Maven repository}'' and published in \emph{The Journal of Systems and Software} in 2017, investigates whether API developers use versioning practices to signal backwards incompatibility, and how unstable interfaces impact client projects in terms of compilation errors~\cite{Raemaekers2017Semantic}.
Although the original study is organized around seven research questions, we decide to focus our effort on three of them that are specifically aimed at understanding the API-client co-evolution problem.
In particular, they address the relationship between \acp{BC} and versioning conventions, and the impact of \acp{BC} on client code.
The main findings of the original study are summarized in the following statements.
Each of these answers one of the research questions we selected:~statement $H_i$ corresponds to question \textbf{Qi}.
In this paper, we  reuse these results as new hypotheses, which we aim to test under different conditions for replication purposes.

\begin{description}
	\item[\rqone] How are semantic versioning principles applied in the Maven Central repository in terms of \acp{BC}?
	\begin{description}
		\item[$H_1$] \textit{\acp{BC} are widespread without regard for semantic versioning principles}.
	\end{description}
	\item[\rqtwo] To what extent has the adherence to semantic versioning principles increased over time?
	\begin{description}
		\item[$H_2$] \textit{The adherence to semantic versioning principles has increased over time}.
	\end{description}
	\item[\rqthree] What is the impact of \acp{BC} on clients? 
	\begin{description}
		\item[$H_3$] \textit{\acp{BC} have a significant impact in terms of compilation errors in client systems}.
	\end{description}
\end{description}

On the one hand, the study relies on \clirr~\cite{Kuhne2003Clirr} to study backwards compatibility.
This tool is used to compute the list of changes between two versions of a Java library.
However, the development of \clirr has stopped in 2005, and \citet{jezek2017api} later showed that it is the least sound of a list of 9 tools for BCs detection in Java.
On the other hand, to identify the impact of \acp{BC} on client code, the original study uses a novel approach that isolates individual changes on the newer version of the API, and injects them one by one in the older version. 
Then, clients are recompiled against each variant of the old version.
The number of compilation errors raised by the Maven compiler is used as a proxy to measure the impact of \acp{BC}.
As the main corpus, the study uses a snapshot of \ac{MCR} dated July 2011, consisting of 148,253 JARs and named the \acl{MDD}.
In the next section, we dive deeper into the design of our replication study to highlight how it differs from the original study in terms of protocol and corpora.

%% file: content/protocol.tex
\section{Design of the Replication Study}
\label{sec:protocol}

\begin{figure}
	\centering
	\includegraphics[width=.9\linewidth]{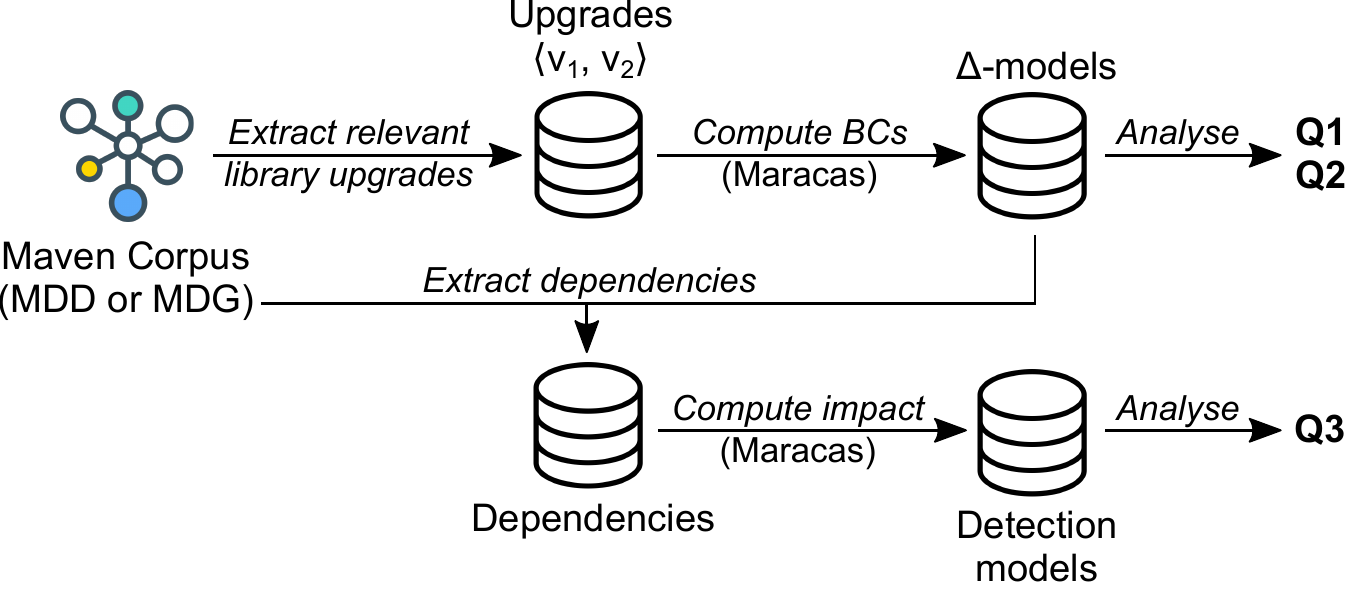}
	\caption{Overview of the analysis protocol.}
	\label{fig:process}
\end{figure}

In this section, we present the protocol of our replication study, summarized in \Cref{fig:process}.
The source material of our study is extracted from two different corpora:~the Maven Dependency Dataset (MDD)~\cite{Raemaekers2013Maven}, which is used in the original study, and the Maven Dependency Graph (MDG)~\cite{benelallam2018mdg}.
These two corpora are snapshots of \ac{MCR} containing metadata information about artefacts, versions, and dependencies between artefacts.
The MDD includes all artefacts from the MCR up to 2011, while the MDG spans seven more years up to 2018.
However, due to subtle differences in the methodology used to build these snapshots, the MDD is not strictly a subset of the MDG.
In this study, we run the very same analysis protocol on both corpora.
On the one hand, re-analysing the MDD enables us to assess the impact of our updated protocol on the results obtained in the original study.
On the other hand, analysing the MDG enables us to broaden the scope of analysed artefacts and strengthen our conclusions.

First, to answer \rqone and \rqtwo, we extract relevant upgrades for all libraries in the corpora, \ie pairs of adjacent releases (\eg JavaServlet versions 4.0.0 and 4.0.1) that conform to the selection criteria presented in \Cref{sec:ds}.
The outputs of this task are the \emph{upgrades datasets} \duo (for the MDD) and \dur (for the MDG).
Then, we use our tool \mrc to compute delta models (\emph{\dmodels}) that store the list of \acp{BC} introduced in a particular upgrade between two versions of a library.
We analyse the resulting \dmodels in \Cref{sec:analysis} to answer our first two research questions.

Second, to uncover the impact of \acp{BC} on client code and answer \rqthree, we build the dependencies datasets \dddo and \dddr which consist of all clients in the corresponding corpus that might be impacted by the changes identified in a \dmodel (\ie all artefacts declaring a dependency towards a library upgrade extracted previously).
We again use \mrc to identify locations in these clients that are impacted by \acp{BC}.
The output is stored in a set of \emph{detection models}, where \acp{BC} are linked to affected locations in client code.
We analyse the resulting models in \Cref{sec:analysis} to answer our last research question.

The remainder of this section is structured as follows.
\Cref{sec:dataset} describes the data extraction process of the protocol.
Then, \Cref{sec:mrc} gives an overview of our static analysis tool, \mrc.
Finally, in \Cref{sec:analysis-approach}, we highlight the key differences between our protocol and the original study's protocol in terms of data selection, filtering, and treatment.

\subsection{Data Extraction}
\label{sec:dataset}
In this section, we introduce the two corpora used in this study, together with the datasets derived from them to answer our research questions.

\subsubsection{Corpora} 
Our study relies on two corpora:~the \acf{MDD} and the \acf{MDG}.
The former is used to verify whether the main findings of the \os hold when following a different protocol, while keeping the same base data.
The latter is included to assess whether the conclusions of the \os remain valid on a larger population, and whether the phenomenon under study (\acp{BC} and \semver in \ac{MCR}) has evolved between 2011 and 2018.

\paragraph{The \acf{MDD}.}
The \ac{MDD} is a publicly available snapshot of the \ac{MCR} dated July 30, 2011~\cite{Raemaekers2013Dataset}.
The corpus contains 148,253 JARs plus additional metadata stored in three different database formats:~MySQL, Berkeley DB, and Neo4j~\cite{Raemaekers2013Maven}.
For our purpose, we rely on the metadata stored in the MySQL database.
More specifically, we consider the \texttt{files} table which stores information about the \pom{groupId}, the \pom{artifactId}, and the \pom{version} of each JAR in the corpus.
There is a minor difference in the number of JARs reported in the original study~\cite{Raemaekers2017Semantic} and the dataset paper~\cite{Raemaekers2013Maven}. 
We consider the information presented in the latter after manually validating the data exposed in the MySQL database.

\paragraph{The \acf{MDG}.}
The \ac{MDG} is a graph-based snapshot of all artefacts on \ac{MCR} as of September 6, 2018~\cite{benelallam2018mdg,benelallam2019maven}. 
It is available as a Neo4j graph database where nodes are Maven artefacts and edges are either dependency relations between two artefacts (denoted \texttt{:DEPENDS}) or upgrade relations between two artefacts of the same library (denoted \texttt{:NEXT}).
The MDG contains 2,4M libraries, 9,7M dependency relations, and 2,1M upgrade relations.
We rely on the \ac{MDG} to extract libraries metadata (\eg versions, clients), and to identify dependency and upgrade relations from which we derive the datasets required for our analyses.

\subsubsection{Derived datasets}
\label{sec:ds}
In what follows, we present the datasets that are derived from each of the two corpora:~the \emph{upgrade datasets} \duo and \dur,  and the \emph{dependencies dataset} \dddo and \dddr.

\paragraph{Upgrades datasets.}
To answer \rqone and \rqtwo, we derive datasets from our corpora consisting of a set of library upgrades $\langle v1 \rightarrow v2 \rangle$.
For our analysis to be accurate and relevant, these library upgrades must fulfil a set of criteria.

As a first filter, we only consider library upgrades $\langle v1 \rightarrow v2 \rangle$ such that $v_1$ and $v_2$ are two versions of the same library (uniquely identified by its \pom{groupId} and \pom{artifactId}) which comply with the \semver scheme.
More precisely, these versions must be of the form \texttt{X.Y[.Z]}, where \texttt{X, Y, Z }$\in \mathbb{N}$, \texttt{X} is the major version, \texttt{Y} the minor version, and \texttt{Z} the (optional) patch version.
Versions suffixed with an additional hyphen-separated qualifier often used to tag release candidates, beta versions, or particular build metadata (\eg \texttt{-b01}, \texttt{-rc1}, \texttt{-beta}, \texttt{-issue101}) are discarded, as they are not meant to be used by the general public.
In the MDG, for instance, we find 328,448 suffixed versions for 8,251 unique suffixes.
The top five most frequent suffixes which we have discarded correspond to release candidates and milestones, namely: \texttt{-rc1} (20,007 occurrences, 6.1\%), \texttt{-rc2} (12,373 occurrences, 3.8\%), \texttt{-M1} (10,614 occurrences, 3.2\%), \texttt{-rc3} (7,829 occurrences, 2.4\%), and  \texttt{-M2} (7,780 occurrences, 2.4\%).
Looking closely at the data, we also noticed that a number of versions, even though they technically comply with \semver, use dates as versions numbers (\eg \texttt{2.5.20110712}).
We decide to discard them, as they do not convey the meaning originally intended by \semver.

Second, $v_1$ and $v_2$ must either be adjacent versions ($v_2$ was released immediately after $v_1$) or separated with non-\semver-compliant versions only (all intermediate versions connecting $v_1$ and $v_2$ through upgrade relations do not match our criteria). For instance, considering the three versions $\langle \texttt{3.0.1} \rightarrow \texttt{3.1-b01} \rightarrow \texttt{3.1.0} \rangle$, only $\langle \texttt{3.0.1} \rightarrow \texttt{3.1.0} \rangle$ would be included.

Third, we only consider upgrades where $v_1$ has at least one external client in the dependency graph (either MDD or MDG).	An \emph{external client} $c$ of a library version $v$ is a Maven artefact such that $c$ depends on $v$ and belongs to a different \pom{groupId}.	This way, we confirm that the artefacts we analyse are indeed used as libraries in practice, and that there are real clients potentially affected by the changes between $v_1$ and $v_2$.
In the MDG, $56\%$ of all artefacts do not have any client ($1,356,413$ out of $2,407,395$), and only $12\%$ of all artefacts ($293,152$) have at least one external client.
In the MDD, $61\%$ of all artefacts do not have any client ($89,772$ out of $148,253$), and only $17\%$ of all artefacts ($24,522$) have at least one external client.

Fourth, as we are only interested in the Java language, we discard all JARs that contain code written in any other JVM-based programming language (\eg Scala, Clojure, Kotlin, Groovy), also hosted on \ac{MCR}. Our heuristic reads the \texttt{source} attribute of \texttt{.class} files, set by most bytecode compilers, to retrieve the source file that was used to produce the bytecode and infer the base language. When languages other than Java are detected in a JAR, it is discarded.

Fifth, to ensure that \mrc can process the JARs accurately, we only select library upgrades for which $v_1$ and $v_2$ are packaged as JAR files and are compiled with a Java version up to 8 included, as the list and semantics of \acp{BC} differ in later versions with the introduction of new language constructs.
This differs from the original study, given that Java 8 was released in 2014 and the original snapshot dates from 2011.
Thus, we expect to report new types of BCs that were not considered for previous Java versions (\eg insertion of a new \ijava{default} method).
The choice of Java 8 is motivated by its popularity: looking at the data, we noticed that it was still by far the most popular Java version on MCR.

Lastly, when looking for clients of libraries, we discard all dependency relations that are not in the \pom{compile} scope or \pom{test} scope since they are either not reliably resolvable (\eg~\pom{provided} and \pom{system} dependencies are not hosted on MCR) or are not included in the compile-time and link-time classpaths of the client and thus cannot impact binary compatibility (\eg~\pom{runtime} dependencies). Only dependencies in the \texttt{compile} and \texttt{test} scopes are considered.

\begin{figure}
	\centering
	\begin{tikzpicture}[artefact/.style={font=\footnotesize}, used/.style={artefact, thick}, unused/.style={artefact, black!50}]
	\node            (v-1)                   {};
	\node [used]     (v0)  [right=.3 of v-1] {3.0.1};
	\node [unused]   (v1)  [right=.3 of v0]  {3.1-b01};
	\node [used]     (v2)  [right=.3 of v1]  {3.1.0};
	\node [unused]   (v3)  [right=.3 of v2]  {4.0.0-b01};
	\node [unused]   (v4)  [right=.3 of v3]  {4.0.0-b02};
	\node [used]     (v5)  [right=.3 of v4]  {4.0.0};
	\node [used]     (v6)  [right=.3 of v5]  {4.0.1};
	\node            (v+1) [right=.3 of v6]  {};
	
	\draw[-{Latex[length=1.5mm,width=1.5mm]}]
	(v0) edge[bend left=1cm] node[above] {$\Delta_{\textsc{Minor}}$} (v2);
	\draw[-{Latex[length=1.5mm,width=1.5mm]}]
	(v2) edge[bend left=.65cm] node[above] {$\Delta_{\textsc{Major}}$} (v5);
	\draw[-{Latex[length=1.5mm,width=1.5mm]}]
	(v5) edge[bend left=1cm] node[above] {$\Delta_{\textsc{Patch}}$} (v6);

	\draw
	(v-1) edge[dotted] (v0);
	\draw
	(v0)  edge[dotted] (v1);
	\draw
	(v1)  edge[dotted] (v2);
	\draw
	(v2)  edge[dotted] (v3);
	\draw
	(v3)  edge[dotted] (v4);
	\draw
	(v4)  edge[dotted] (v5);
	\draw
	(v5)  edge[dotted] (v6);
	\draw
	(v6)  edge[dotted] (v+1);
	\end{tikzpicture}
	\caption{Extracting relevant upgrades from the JavaServlet project (\texttt{javax.servlet :javax.servlet-api}) between versions \texttt{3.0.1} and \texttt{4.0.1}. Dotted lines denote upgrade relationships between Maven artefacts. Only major, minor, and patch upgrades are analysed:~release candidates, alpha and beta versions, and other qualified versions are discarded. Here, \dmodels are computed for the upgrades $\langle 3.0.1 \rightarrow 3.1.0 \rangle$, $\langle 3.1.0 \rightarrow 4.0.0 \rangle$, and $\langle 4.0.0 \rightarrow 4.0.1 \rangle$.}
	\label{fig:javax-servlet}
\end{figure}
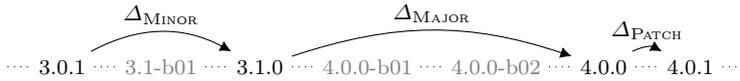

As an illustration of the selection process, \Cref{fig:javax-servlet} depicts how interesting upgrades are picked up between versions \texttt{3.0.1} and \texttt{4.0.1} of the JavaServlet library, and how \dmodels are classified as major, minor, or patch.

From the original corpus (MDD), we obtain the upgrades dataset \duo consisting of $11,384$ upgrade pairs, along with the associated \dmodels computed using \mrc.
This dataset differs from the one presented in the original study which contains $126,070$ pairs~\cite{Raemaekers2017Semantic}.
This difference is explained by the additional filters employed in our protocol:~most upgrades are discarded because they do not have any external client;~$848$ because they contain bytecode generated from other JVM-based languages ($2$ in Clojure, $71$ in Groovy, $76$ in Scala, $699$ a mix of these or other languages);~$641$ because of an invalid Java version; $306$ because the JARs could not be retrieved from MCR;~$2$ because \mrc raised an exception when processing the JARs;~$31$ because they use dates as versions;~$309$ because the metadata states that $v_1$ of the library was released after $v_2$, and;~$27$ that have an erroneous release date strictly greater than 2011. 

From the replication corpus (MDG), we obtain the upgrades dataset \dur consisting of $119,879$ upgrade pairs.
Most upgrades are discarded because they do not have any external client;~$39,986$ because they are written in other JVM-based languages ($20$ in Clojure, $1,137$ in Groovy, $1,359$ in Kotlin, $14,402$ in Scala, $23,068$ a mix of these or other languages);~$852$ because of an invalid Java version; $10,588$ because the JARs could not be retrieved from Maven Central;~$271$ because \mrc raised an exception when processing the JARs;~$115$ because they use dates as versions, and;~$2,929$ because the metadata states that $v_1$ of the library was released after $v_2$.

\begin{figure}[bt]
	\centering
	\begin{subfigure}[b]{.49\textwidth}
		\centering
		\includegraphics[width=\textwidth]{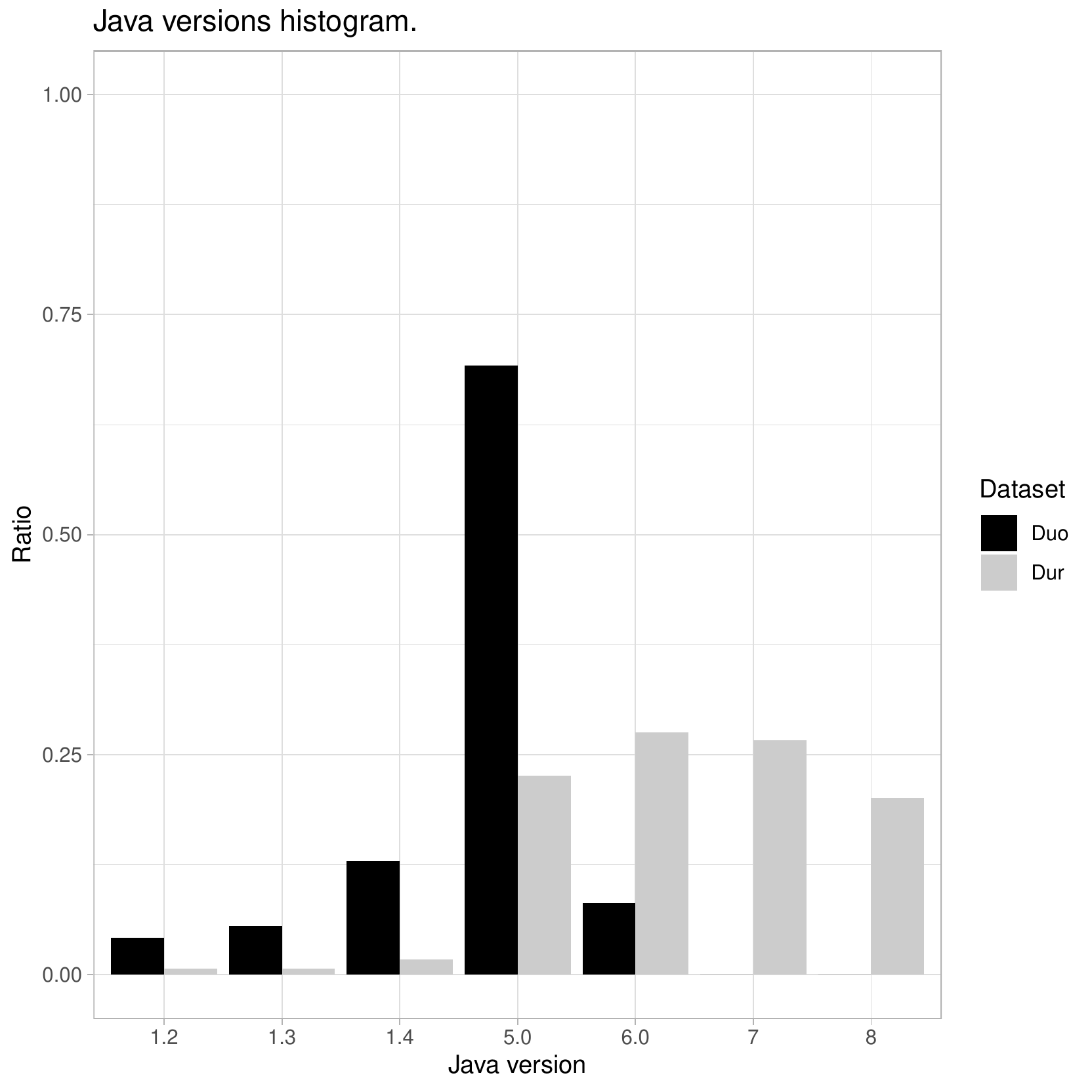}
	\end{subfigure}
	\hfill
	\begin{subfigure}[b]{.49\textwidth}
		\centering
		\includegraphics[width=\textwidth]{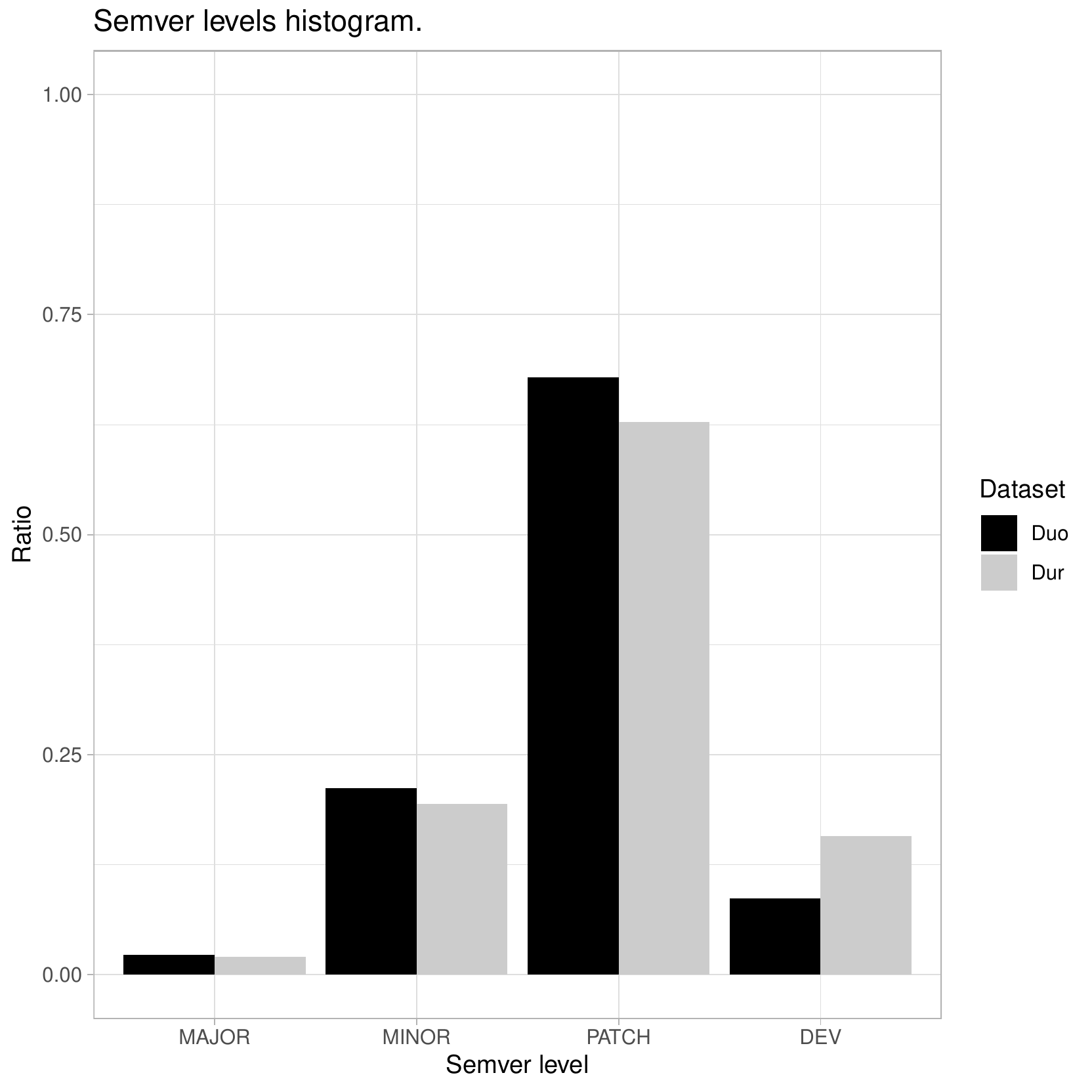}
	\end{subfigure}
	\caption{Java versions and \semver levels histograms.}
	\label{fig:upgrades-histos}
\end{figure}

\Cref{tab:upgrades-stats} and \Cref{fig:upgrades-histos} summarize some descriptive statistics of both datasets.
As most distributions are strongly skewed and difficult to visualize (number of clients, size, \etc), \Cref{tab:upgrades-stats} lists their minimum, maximum, median, mean, and quartile values.
As an illustration, the top five most popular libraries in \dur are \texttt{commons-io 2.4} (36,186 clients), \texttt{slf4j-api 1.7.21} (33,582 clients), \texttt{commons-codec 1.10} (32,990 clients), \texttt{slf4j-api 1.7.12} (25,317 clients), and \texttt{slf4j-api 1.7.7} (22,939 clients).
We refer the reader to the companion webpage and Zenodo repository\footnote{\url{https://zenodo.org/record/5221840}} to access and interact with the datasets.

\begin{table}[bt]
	\centering
	\caption{Descriptive statistics of the datasets \duo and \dur.}
	\label{tab:upgrades-stats}
	\begin{tabular}{lrrrrrr}
	Dimension & Min. & Q1 & Median & Mean & Q3 & Max. \\
	\midrule
	\multicolumn{7}{c}{\duo} \\
	\midrule
 	External Clients & 1  & 1 & 3 & 24.31 & 8 & 6,153 \\
	\# of releases & 1 & 1 & 1 & 1.23 & 1 & 55 \\
	Age (in months) & 1 & 1 & 2 & 5.55 & 5 & 146 \\
	Releases / month & 0.01 & 0.25 & 0.5 & 0.53 & 1 & 18 \\
	\# of decls. & 1 & 71 & 280 & 2,076 & 1,276 & 218,274 \\
	\# of API decls. & 1 & 49 & 200.5 & 1,515.6 & 891.2 & 159,478 \\
	\midrule
	\multicolumn{7}{c}{\dur} \\
	\midrule
	External Clients & 1 & 1 & 2 & 25.31 & 7 & 36,186 \\
	\# of releases & 1 & 7 & 20 & 39.4 & 49 & 587  \\
	Age (in months) & 1 & 10 & 24 & 34.38 & 49 & 158 \\
	Releases / month & 0.01 & 0.4 & 0.85 & 1.96 & 1.79 & 320 \\
	\# of decls. & 1 & 79 & 329 & 2,519 & 1,346 & 586,172 \\
	\# of API decls. & 0 & 52 & 220 & 1,850 & 974 & 561,465 \\
	\end{tabular}
\end{table}

\paragraph{Dependencies datasets.}
To answer \rqthree, for each upgrade in \duo and \dur, we compute the list of all clients potentially impacted.
That is, all clients that declare a compile-time or test-time dependency towards the library $v1$ in a $\langle v1 \rightarrow v2 \rangle$ upgrade pair, which are potentially affected by the \dmodel between $v1$ and $v2$.
We observe that, often, different versions of the same client (\eg $c_{v_1}$, $c_{v_2}$, and $c_{v_3}$) all depend on the same library version $v1$.
In such a case, it is unlikely that $c_{v_1}$ would migrate to $v_2$ as it is superseded by $c_{v_3}$.
Thus, we only include $c_{v_3}$ in the resulting datasets to avoid counting the impact of the \dmodel between $v1$ and $v2$ on $c$ multiple times.
The resulting datasets for \duo and \dur are \dddo and \dddr which contain, respectively, 35,539 and 293,817 clients.

\subsection{Maracas}
\label{sec:mrc}
\mrc is a new static analysis tool written in Rascal~\cite{rascal} and Java, which allows us to (i)~automatically compute a \dmodel between two binary versions of a library and (ii)~detect locations in a client binary that are affected by the BCs listed in a \dmodel.

\subsubsection{\dmodels.}
A \dmodel is a model that stores the list of \acp{BC} between two versions $\langle v_1 \rightarrow v_2 \rangle$ of a library.
To compute the \dmodel between two versions of a library, \mrc internally relies on \japi.
\japi is a tool that compares two JAR files and generates a list of \acp{BC} between these two JARs.
It is able to identify 31 binary incompatible BCs as specified in the \ac{JLS} 8$^{th}$ Edition~\cite{Gosling2014Java}.
Example of BCs include removals (\eg \texttt{fieldRemoved}, \texttt{methodRemoved}), changes in modifiers (\eg \texttt{methodNowAbstract}, \texttt{classNowFinal}) and visibilities (\eg \texttt{fieldLessAccessible}), type changes (\eg \texttt{methodReturnTypeChanged}), to name a few.\footnote{A complete list and description of these BCs is available on the companion webpage (\url{https://crossminer.github.io/maracas/emse21}) and Zenodo repository (\url{https://zenodo.org/record/5221840})}
As we shall see later in \Cref{sec:rq4-protocol}, some are more critical than others in terms of impact on clients.
A \dmodel in \japi follows a tree structure and consists of a list of modified types (classes, interfaces) that recursively contain all modified child elements (\eg methods, fields, modifiers).
Modified elements themselves are labelled with a kind of \ac{BC} (\eg \texttt{classRemoved}, \texttt{fieldNowFinal}).
\mrc transforms \japi's tree-structured models into a value in Rascal conforming to a \dmodel \ac{ADT}, which we use for further analysis.
The choice of \japi is motivated by its high popularity and accuracy~\cite{jezek2017api}, and by its active community.

Atop \japi, we implement in \mrc a mechanism to classify declarations in a library as \emph{stable} or \emph{unstable}.
Unstable declarations are code elements (\eg classes, methods, fields) that are explicitly marked with a specific annotation or that are contained in a package not meant to be used by clients.
For instance, Google libraries use the annotation \texttt{@Beta} and Apache libraries use the annotations \texttt{@Internal} to denote code elements that are subject to sudden changes or that should not be used by clients.
Similarly, Eclipse packages containing the word \texttt{internal} and JDK packages starting with \texttt{sun.*} should not be considered part of the API~\cite{businge2019stable}.
Because declarations that are explicitly marked as unstable by developers escape the rules of \semver, it is important to classify stable and unstable declarations.
To come up with a list of annotations to consider, we automatically extracted all annotations used in the 100 most popular libraries on \ac{MCR} and manually reviewed their documentation to state whether they are used to delimit unstable APIs.
This way, we extracted 185 unique annotations and a list of keywords that typically appear in their name (\textit{api, alpha, beta, internal, protected, private, restricted, experimental, dev, access}).
Then, we conducted a keyword-based search on the annotations used in the top 1,000 most popular libraries on \ac{MCR} using as input the keywords extracted manually.
This way, we extracted $1,258$ annotations of which 48 matched a keyword.
The five most common API annotations encountered in these 1,000 libraries are as follows: \texttt{@Beta} (1,451 occurrences), \texttt{@InterfaceAudience} (1,819 occurrences), \texttt{@InternalApi} (1,414 occurrences), \texttt{@Internal} (716 occurrences), and \texttt{@SdkInternalApi} (607 occurrences).
In Maracas, we use the list of extracted keywords to extract package and annotation names used to delimit unstable APIs.
Then, we classify each BC in the \dmodels according to whether they affect a stable or unstable declaration of the library.

\subsubsection{Detection models.}
Prior work uses two main techniques to assess the impact of \acp{BC} on client projects:~either by tracing library types that are imported in client code (through \texttt{import} statements in Java)~\cite{Bavota2013Evolution, Xavier2017Historical} or by measuring the ripple effect of changes on clients that have already been migrated manually by developers~\cite{Robbes2012How}.
The former approach largely over-estimates the impact of \acp{BC} (a client may not use the broken declaration in the imported type) and the latter requires the availability of migrated clients.
The \os employs a novel technique that consists in isolating and injecting each individual \acs{BC} in the old library's source code.
Then, every client is compiled against every ad-hoc version of the library where a single \acs{BC} is inserted to measure its impact in terms of compilation errors~\cite{Raemaekers2017Semantic}.
To the best of our knowledge, however, it is rarely possible to inject individual changes in a library without having to refactor other parts of the API.
Removing or renaming a method, for instance, triggers a ripple effect within the library itself, which results in multiple changes being inserted and impacting the clients.
Therefore, we hypothesize that this technique overestimates the impact of \acp{BC} on clients.
Moreover, while measuring the impact of \emph{source incompatible} BCs by counting compilation errors is a valid approach, it is not appropriate to measure the impact of \emph{binary incompatible} BCs which are instead checked by the JVM linker.
This motivates the need of having a dedicated tool for identifying \acp{BC} impact using static analysis.

\mrc leverages the \dmodels and Rascal \mt models to link \acp{BC} to affected client declarations using static analysis of binary code.
An \mt model is an \ac{ADT} that models relations between Java elements, extracted from a JAR file, in immutable binary relations (\eg containment relations among classes and method declarations, invocation relations among method declarations)~\cite{Basten2015M3}.
Internally, \mt relies on the ASM framework\footnote{\url{https://asm.ow2.io}} to parse Java bytecode and populate the relations.
By combining information about breaking declarations in \dmodels and uses of these declarations in client code using \mt model, \mrc is able to mark affected client declarations. 
This detection algorithm is based on the \ac{JLS} specification~\cite{Gosling2014Java}, and its implementation in \mrc for each kind of \ac{BC} is detailed on the companion webpage.
The output of this task is a set of \emph{detections} that point to the affected client element, the modified API element, the way it is being used (\eg \texttt{methodInvocation}, \texttt{fieldAccess}, \texttt{implements}), and the type of \ac{BC}.
In the example of \Cref{lst:bc-example}, the affected client element is \texttt{MockHttpServletRequest} which uses the modified API element \texttt{changeSessionId()} through an \texttt{implements} relation due to a \texttt{methodAddedToInterface} change.

\subsubsection{Limitations}

\paragraph{Overridden methods.}
	\mt models generated from Java binaries do not have information related to overridden methods. 
	This means that \mrc cannot detect \acp{BC} impact on code that uses the API through method overriding.
	For instance, the \emph{method now final} BC breaks clients that override the now-final method, but has no impact on clients that do not override this method.
	In such a case, due to the lack of information, we follow a pessimistic approach and report a detection in all cases.
\paragraph{Exceptions handling.}
	Information related to thrown and caught exceptions is not part of the \mt models.
	\mrc has no information related to the types of exceptions handled in the \ijava{try-catch} statements of clients.
	Thus, if a method in a library throws a new kind of checked exception, \mrc is not able to state whether the client will be impacted.
	In this case, we follow a pessimistic approach and always report a detection.

\paragraph{Inheritance hierarchy.}
	Changing the type of a field, method, parameter, or any other member, or casting might turn out to be a generalization or specialization of the associated type.
	A type is generalized when it is changed to a supertype and a type is specialized when it is changed to a subtype.
	In \mrc we only have access to the client and to the analysed API binaries.
	Other APIs used by the client are not part of the analysis. 
	Therefore, when a type is changed in a library, we cannot build the whole inheritance hierarchy to state whether this type change corresponds to a generalization or specialization.
	Without this information, \mrc might report false positive detections, following a pessimistic approach.

\paragraph{The super keyword.}
	The \ijava{super} keyword in Java gets a special treatment when detecting errors caused by \acp{BC} at the client level.
	When the visibility of a constructor goes from public to protected, and the constructor is invoked through the use of the \ijava{super} keyword in the subtype constructor, no error should be reported. 
	However, if the constructor is invoked without using the \ijava{super} keyword, an error should be reported.
	\mrc is not able to differentiate between these two types of invocations, and thus follows a pessimistic approach to always report a detection.

\paragraph{The strictfp and native modifiers.}
	\japi does not report \acp{BC} related to the \ijava{strictfp} and \ijava{native} modifiers.
	Therefore, \mrc is unable to detect client code affected by changes related to these modifiers, directly affecting the recall of the tool.

\subsubsection{Validation.}
\mrc is the cornerstone tool of our approach as it is used to both compute the \dmodels revealing BCs (using \japi under the hood) and the detection models revealing their impact.
To correctly interpret our results, it is essential to analyse the accuracy of \mrc.

When \mrc cannot accurately state whether a BC actually has an impact due to the limitations listed above, the approach we follow is to always over-approximate the detections at the cost of sometimes reporting false positives, while avoiding any false negative.
This means that the results we obtain regarding BCs and their impact might be slightly overestimated, but they are not underestimated:~every BC and detection we report does exist.

Binary compatibility is checked by the JVM linker.
To evaluate \mrc, we aim to compute its accuracy by comparing \mrc detections with the error messages thrown by the JVM linker itself when encountering code impacted by BCs.
The JLS states that binary compatibility should be checked during the loading and linking phases of the JVM, but the choice of implementing lazy or eager initialization of classes is up to the implementors.
In practice, the reference implementation (OpenJDK HotSpot) implements lazy loading and waits for class initialization to load and link a class.
It follows that, to record the errors thrown by the linker, it is necessary to \emph{execute} a Java program making use of breaking declarations.
In addition, the Java linker throws an exception and stops processing the class after the first error is encountered, so the executed Java programs should contain only a single use of a breaking declaration to record all errors.

To evaluate the accuracy of \mrc, we thus reuse and extend the benchmark proposed by \citeauthor{jezek2017api}~\cite{jezek2017api}.
Their benchmark consists of a library $v1$, a library $v2$ that breaks $v1$ in all possible ways, and a client $c$ that uses all declarations of $v1$ in various ways.
In order to trigger linking errors, client $c$ consists of a set of \texttt{Main} files.
In the original version of the benchmark, however, some \texttt{Main} files use several declarations of the library $v1$:~if the first use fails, the others are not evaluated by the linker.
Thus, we split the client $c$ into more cases so that every case exercises a single declaration of $v1$.

Our final benchmark for detections consists of 345 cases, where each case consists of a Java entry point (\texttt{Main} file and method) that exercises one particular BC and one particular way of using it.
The benchmark script first compiles $v1$, $v2$, and $c$ in their binary form (JAR), and then attempts to run every single \texttt{Main} file in $c$, replacing $v1$ with $v2$ in its classpath.
Whenever a linking error is encountered, it is written to disk.
Then, we run \mrc giving it $v1$, $v2$, and $c$ as inputs to get the list of detections.
If a detection matches an error reported by the linker, it is a true positive, if it does not match any linker error it is a false positive, and if there is no detection for a particular linker error it is a false negative.
To support future research, we have made our benchmark publicly available on the companion webpage.

Out of the 345 cases, the JVM linker reports 132 errors and \mrc reports 135 detections.
Out of the 135 detections, 130 are true positives and 5 are false positives.
There are 2 false negatives.
In this benchmark, \mrc achieves a precision of 96.3\% and a recall of 98.5\%.
The five false positives are due to the limitation listed in the section above.
The two false negatives are due to a limitation of \japi, which does not compute BCs related to the \ijava{strictfp} and \ijava{native} modifiers.

In addition to this benchmark, we developed a test suite as part of \mrc consisting of 402 test cases.
Using our own test cases, we highlighted a bug in \japi which we fixed through a pull request accepted by the project maintainers.\footnote{\url{https://github.com/siom79/japicmp/pull/251}}

\subsection{Analysis Approach}
\label{sec:analysis-approach}
In this section, we compile some of the most relevant aspects of the analysis performed in this study, and we contrast them against the original study (\cf \Cref{tab:ds}).
We refer the reader to Section 4 of the original work for further information.

\begin{table*}[t]
	\centering
	\caption{Main commonalities and differences between the original study and the replication study protocols.}
	\label{tab:ds}
	\begin{tabular}{llp{4.1cm}}
		 & Original Study & Replication Study \\
		\midrule
		Corpus & MDD~\cite{Raemaekers2013Maven} & MDD+MDG~\cite{benelallam2018mdg} \\
		Corpus date interval & 2005--2011 & 2005--2018 \\
		Backwards compatibility type & Binary & Binary \\
		Backwards compatibility tool & \texttt{clirr} & \japi  \\
		Compared versions & Adjacent & Adjacent \\ 
		Versioning scheme & \semver & \semver \\
		Languages & JVM-based & Java \\
		Clients per library & $\geq 0$ & $\geq 1$ \\ 
		Client impact detection & Compilation errors & Static analysis \\ 
		Code-level mechanisms & \texttt{@Deprecated} & Annotations, package naming
	\end{tabular}
\end{table*}

\paragraph{Backwards compatibility.}
The \os computes binary incompatible changes with \clirr.
However, \clirr is not able to report \acp{BC} related to exceptions and generics and misinterprets changes related to inheritance and other modifiers~\cite{jezek2017api}.
In this study, we use \japi to compute both source and binary incompatible changes.
The latter performs better than \clirr according to \citet{jezek2017api}.
Although \japi is unable to identify changes related to generics---which, due to type erasure in Java, does not impact binary compatibility analysis---it accurately reports all changes related to exceptions and inheritance.
In addition, it is more accurate than \clirr when reporting on changes associated with modifiers.
We also contributed to the tool by fixing a bug related to the detection of modifier changes. 
Other committers have also made some contributions to improve \japi accuracy in recent times.
With these changes, one new case within the \citet{jezek2017api}'s benchmark passes: the decrease of a nested interface access modifier from \ijava{public} to \ijava{protected}. 

\paragraph{Library and version selection.}
As is the case in the original study, we only compute deltas between adjacent versions of an API, which strictly follow the \texttt{X.Y[.Z]} version convention.
However, in contrast, we only consider artefacts that have at least one external client on the \ac{MCR}.
This way, we ensure that the artefacts we analyse are indeed used as libraries by clients.
It is a significant difference with the original study, as only $17\%$ of all artefacts in the MDD and $12\%$ of all artefacts in the MDG have at least one external client.
We also account for initial development releases (\texttt{0.Y[.Z]}) which are not considered in the original study.

\paragraph{Parallel branches and maintenance releases.}
The \os does not account for maintenance releases that happen in practice.
For instance, suppose version \texttt{2.4} is released after \texttt{3.0}, as a maintenance release for the \texttt{2.X} branch.
Using release dates to infer the order of versions, \acp{BC} for the upgrade $\langle \texttt{3.0} \rightarrow \texttt{2.4} \rangle$ would be computed, even though this is not the expected behaviour.
Instead, we employ the MDG which properly represents these upgrade relations regardless of release date, and accounts for maintenance releases.

\paragraph{Breaking changes impact.}
The \os detects client code affected by \acp{BC} by means of injecting changes in the source code of an API.
 After the code injection, the client is compiled against the modified API and new compilation errors are recorded.
 There might be pre-existing compilation errors before changes are injected in the API. 
 These errors are intentionally excluded from the analysis.
 However, this approach introduces a set of limitations that can affect the outcome of the study, some of them have already been identified by \citet{Raemaekers2017Semantic}.
 We describe them as follows:
 \begin{enumerate*}[label=(\roman*)]
 	\item injecting changes in isolation might introduce compilation errors that must be fixed. In some cases, multiple changes should be injected at the same time in order to avoid introducing compilation errors;
 	
 	\item pre-existing errors might hide new errors related to the injected \ac{BC};
 	
 	\item reporting on compilation errors gives us an idea of how source compatibility is affected. However, binary compatibility is not equivalent to source compatibility. Compilation errors account for source incompatible changes and cases of binary incompatibility that are shared between both sets;
 	
 	\item we cannot guarantee that all expected compilation errors are reported by the compiler. For instance, if at least one imported package in a class cannot be found, the compiler will not reach subsequent errors~\cite{Raemaekers2017Semantic};
 	
 	\item when injecting changes in the API, it is difficult to manage cases where one piece of code is related to multiple \acp{BC}. For instance, we inject a piece of code related to changes $C_1$ and $C_2$ in a given API.  If we want to measure the impact of both changes, we will end up with the same number of compilation errors for both cases without discriminating their origin, and;
 	
 	\item the compiler cannot tell the cause or the \ac{BC} that produces a given error.
 \end{enumerate*}

Overall, given the widespread use of the language features involved in the above possible causes of inaccuracy, and their relation to the research questions, we believe that developing and using a more accurate tool will have a significant impact on the outcome.
Thus, we use \mrc to detect affected code on the client side and report on its accuracy.

 \paragraph{Deprecated and unstable interfaces.}
 The \os uses \texttt{@Deprecated} annotations to identify unstable interfaces.
 Occurrences of this annotation are computed, except for nested cases.
 This means that the analysis will not detect declarations within a deprecated class, where explicit annotations have not been used~\cite{Raemaekers2017Semantic}.
 These cases are considered in the present work.
 Moreover, there are also other mechanisms to signal unstable interfaces.
 We argue that other annotations, such as Google's \texttt{@Beta} and Apache's \texttt{@Internal} annotations, are also used to signal instability in an API. 
 In addition, naming conventions on packages are also used for the same purpose. 
 We then include the detection of these cases to perform a deeper analysis of the derived datasets.

%% file: content/analysis.tex
\section{Results \& Analysis}
\label{sec:analysis}

In this section, we analyse the data extracted using the protocol described in~\Cref{sec:protocol}. 
Each subsection describes the method, results, and analysis of a particular research question.

\subsection{\rqone: How are semantic versioning principles applied in the Maven repository in terms of \acp{BC}?}
\label{sec:rq1-protocol}

\subsubsection{Method}
With \rqone, we analyze when and where BCs happen and with which frequency.
We attempt to distinguish expected and unexpected BCs according to the \semver principles and the use of code-level mechanisms to signal unstable APIs.
In a first step, we seek to highlight the impact of the updated protocol described in~\Cref{sec:protocol} on the results reported in the original study.
To do so, we compute the \dmodels between every $\langle v_1 \rightarrow v_2 \rangle \in \mathcal{D}_u^o$, while distinguishing among major, minor, patch, and initial development releases.
In a second step, to assess whether the results hold on the larger dataset \dur comprising seven more years of Maven Central, we run the same analysis for every $\langle v_1 \rightarrow v_2 \rangle \in \mathcal{D}_u^r$.
The \dmodels distinguish between BCs that are introduced in stable and unstable parts of the APIs, according to code-level annotations (such as \texttt{@Beta} or \texttt{@Internal}, \cf \Cref{sec:protocol}), as well as naming conventions (such as \emph{internal}).
As BCs in unstable parts of an API are to be expected, only BCs introduced in the stable parts are included.

To know where to expect BCs or in which type of upgrades, we compare the percentage of breaking upgrades per \semver level.
Alongside \semver categories (\ie major, minor, patch, and initial development), we also consider the group of non-major releases as a whole (\ie~minor and patch releases combined).
Then, to know how many BCs are usually introduced in each \semver level, we consider the distribution of BCs over all groups.

We wrap up the analysis of \rqone by studying the frequency of each type of BC.
From these results, we identify the most common BCs in our datasets and compare these results against the ones presented in the \os.

\subsubsection{Results}
\label{sec:res-rq1}

\paragraph{Breaking upgrades.}
\Cref{tab:rq1} highlights the main results obtained for \rqone.
The first block lists the results reported in the \os~\cite{Raemaekers2017Semantic}, the second block the results obtained for \duo (for replication purposes), and the third block the results obtained for \dur.

First of all, we compare the results reported in the original study against those obtained for \duo.
While we report a similar number of breaking upgrades overall (\cf \emph{Total} row:~32.2\% in \duo vs. 30.0\% in the original study), we observe that the difference in the ratio of breaking upgrades per \semver level is stronger (\cf \emph{Major}, \emph{Minor}, \emph{Patch}, and \emph{Initial development} rows).
As expected, most major upgrades in \duo introduce \acp{BC} (72.7\%), which contrasts with the results obtained in the \os (35.9\%).
While the \os reports that there are as many breaking major upgrades as breaking minor upgrades, we observe a sharper difference between these two levels in the same corpus:~72.7\% of major upgrades (vs. 35.9\% in the original study) and 50.1\% minor upgrades (vs. 35.7\% in the original study) break in \duo.
With regards to patch upgrades, we observe a similar percentage of breaking cases (24.2\% in \duo vs. 23.8\% in the \os).
39.3\% of the initial development upgrades, which are not considered in the original study, are breaking.
Overall, we report that 30.5\% of non-major releases do not conform to \semver, which matches the results obtained in the \os (29.0\%).
Upgrades that comply with the scheme principles (\ie major upgrades, initial development upgrades, and non-breaking minor and patch upgrades) represent 72.8\% of all upgrades in \duo.

For \dur, which spans seven more years of the \ac{MCR} and comprises ten times more upgrades, we observe that the tendency to comply with \semver improves.
The ratio of breaking upgrades is lower overall (22.0\% in \dur vs. 32.2\% in \duo), and for each level: 61.8\% for major upgrades, 37.9\% for minor upgrades, 14.6\% for patch upgrades, and 26.7\% for initial development upgrades.
This amounts to 83.4\% of all upgrades conforming to \semver principles.
However, 20.1\% of non-major upgrades are still breaking and thus do not comply with the versioning conventions.
The difference in results between \duo and \dur suggests that the adherence to semantic versioning may have increased over time.
This intuition is investigated further in the next research question \rqtwo.

\begin{table}[tb]
	\centering
	\caption{Total and breaking upgrades in the original study, \duo, and  \dur datasets.}
	\begin{tabular}{lrrrr}
		& \multicolumn{2}{c}{Total} & \multicolumn{2}{c}{Breaking} \\
		\cmidrule{2-5}
		Level & Count & \% & Count & \% \\
		\midrule
		\multicolumn{5}{c}{Original study} \\
		\midrule
		Major     & 11,892 & 14.8 & 4,268  & 35.9 \\
		Minor     & 29,957 & 37.2 & 10,690 & 35.7 \\
		Patch     & 38,740 & 48.1 & 9,239  & 23.8 \\
		Dev  & n/a    & n/a  & n/a    & n/a  \\
		Non-major & 68,697 & 85.3 & 19,929 & 29.0 \\
		Total     & 80,589 & 100  & 24,197 & 30.0 \\
		\midrule
		\multicolumn{5}{c}{\duo} \\
		\midrule
		Major     & 253    & 2.2    & 184    & 72.7 \\
		Minor     & 2,413  & 21.2  & 1,228 & 50.1 \\
		Patch     & 7,728 & 67.9 & 1,870 & 24.2 \\
		Dev  & 990  & 8.7  & 389 & 39.3 \\
		Non-major & 10,141 & 89.1 & 3,098   & 30.5 \\
		Total     & 11,384 & 100  & 3,671 & 32.2 \\
		\midrule
		\multicolumn{5}{c}{\dur} \\
		\midrule
		Major     & 2,431   & 2.0  & 1,503  & 61.8 \\
		Minor     & 23,309  & 19.4 & 8,837 & 37.9 \\
		Patch     & 75,282  & 62.8 & 11,031 & 14.6 \\
		Dev & 18,857  & 15.7 & 5,036  & 26.7 \\
		Non-major & 98,591 & 82.2 & 19,868 & 20.1 \\
		Total     & 119,879 & 100 & 26,407 & 22.0
	\end{tabular}
	\label{tab:rq1}
\end{table}

\paragraph{Number of BCs.}

\begin{figure}[tb]
	\centering
	\begin{subfigure}[b]{.49\textwidth}
		\centering
		\includegraphics[width=\textwidth]{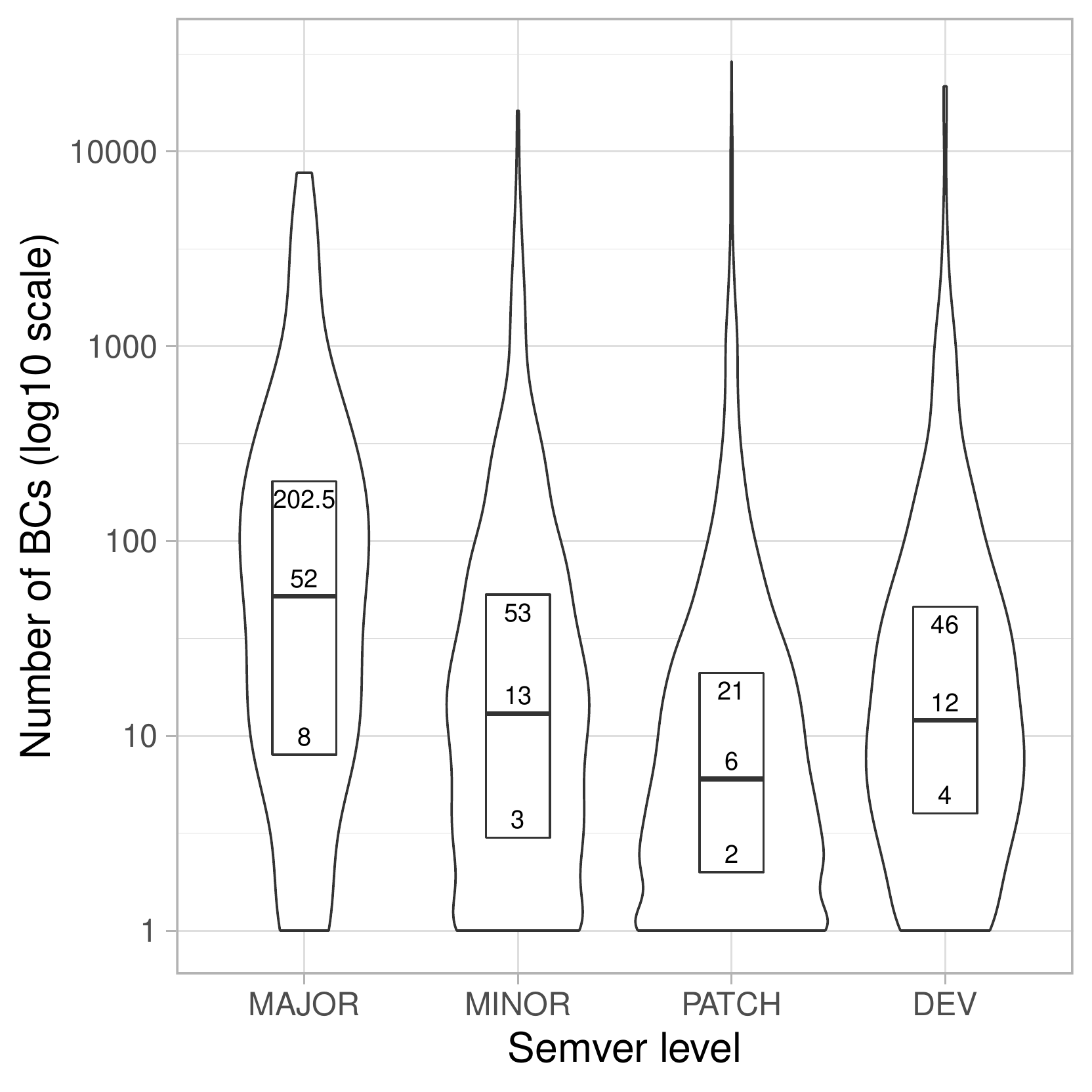}
		\caption{\duo}
	\end{subfigure}
	\hfill
	\begin{subfigure}[b]{.49\textwidth}
		\centering
		\includegraphics[width=\textwidth]{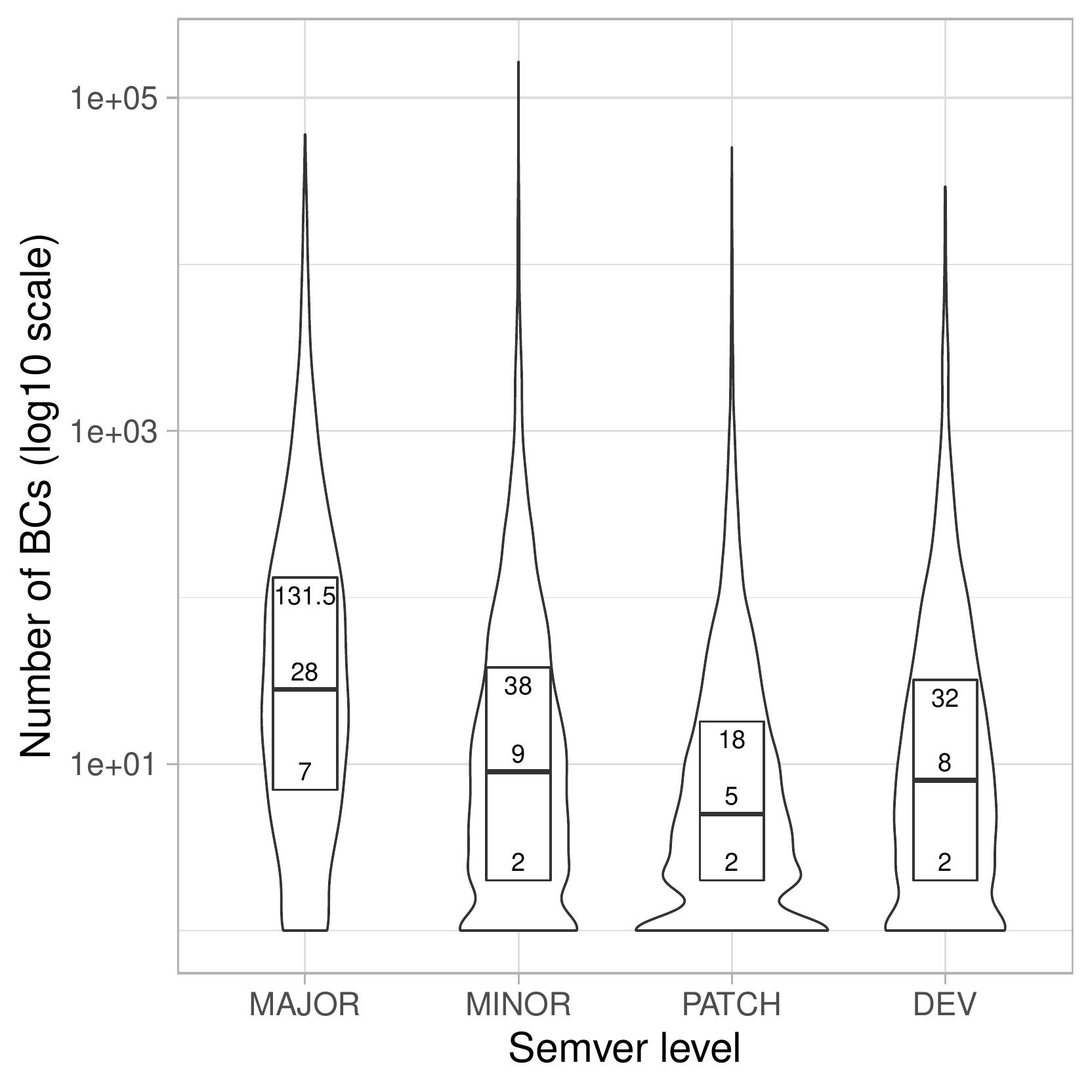}
		\caption{\dur}
	\end{subfigure}
	\caption{Violin plots of the number of BCs in breaking upgrades per \semver level. Inside each violin plot, we display the first quartile, the median and the third quartile.}
	\label{fig:bc-number}
\end{figure}

To study the frequency of BCs introduction, we first look at the number of BCs introduced in breaking releases, \ie releases that contain at least one BC.
\Cref{fig:bc-number} shows the distribution in a logarithmic scale of BCs per \semver level for breaking releases in \duo and \dur.
Looking at the median values, we notice that the number of BCs is higher in major upgrades (52 in \duo and 28 in \dur).
Minor and initial development upgrades tend to have a similar number of BCs in both datasets (13 and 12 in \duo, and 9 and 8 in \dur, respectively).
Patch upgrades introduce the least number of BCs (6 in \duo and 5 in \dur).
This suggests that non-major development releases not only do break less often, they also tend to introduce fewer BCs when they do.

\begin{figure}
	\centering
	\begin{subfigure}[t]{0.8\hsize}
		\centering
		\includegraphics[width=\hsize]{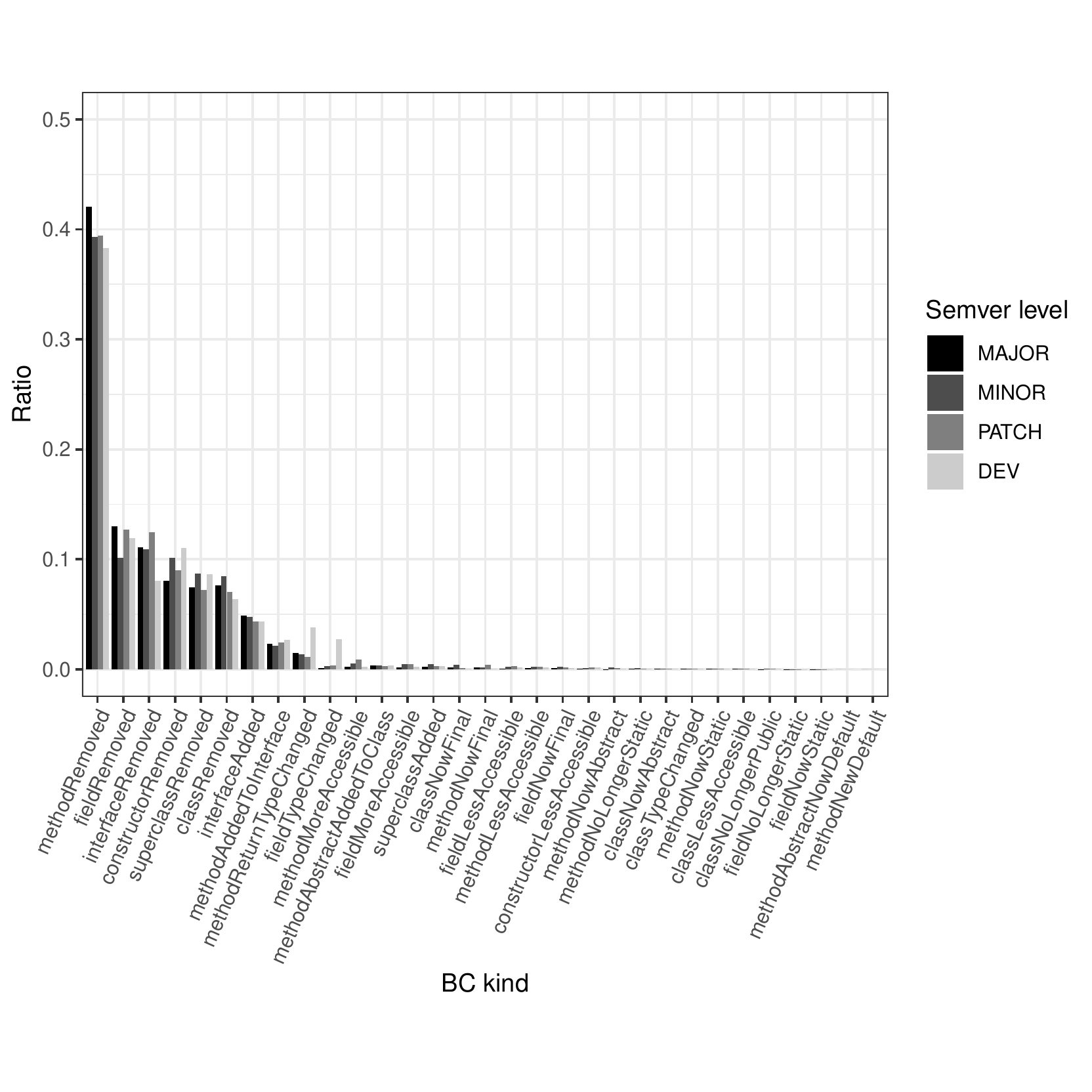}
		\caption{\duo}
	\end{subfigure}
	\hfill
	\begin{subfigure}[t]{0.8\hsize}
		\centering
		\includegraphics[width=\hsize]{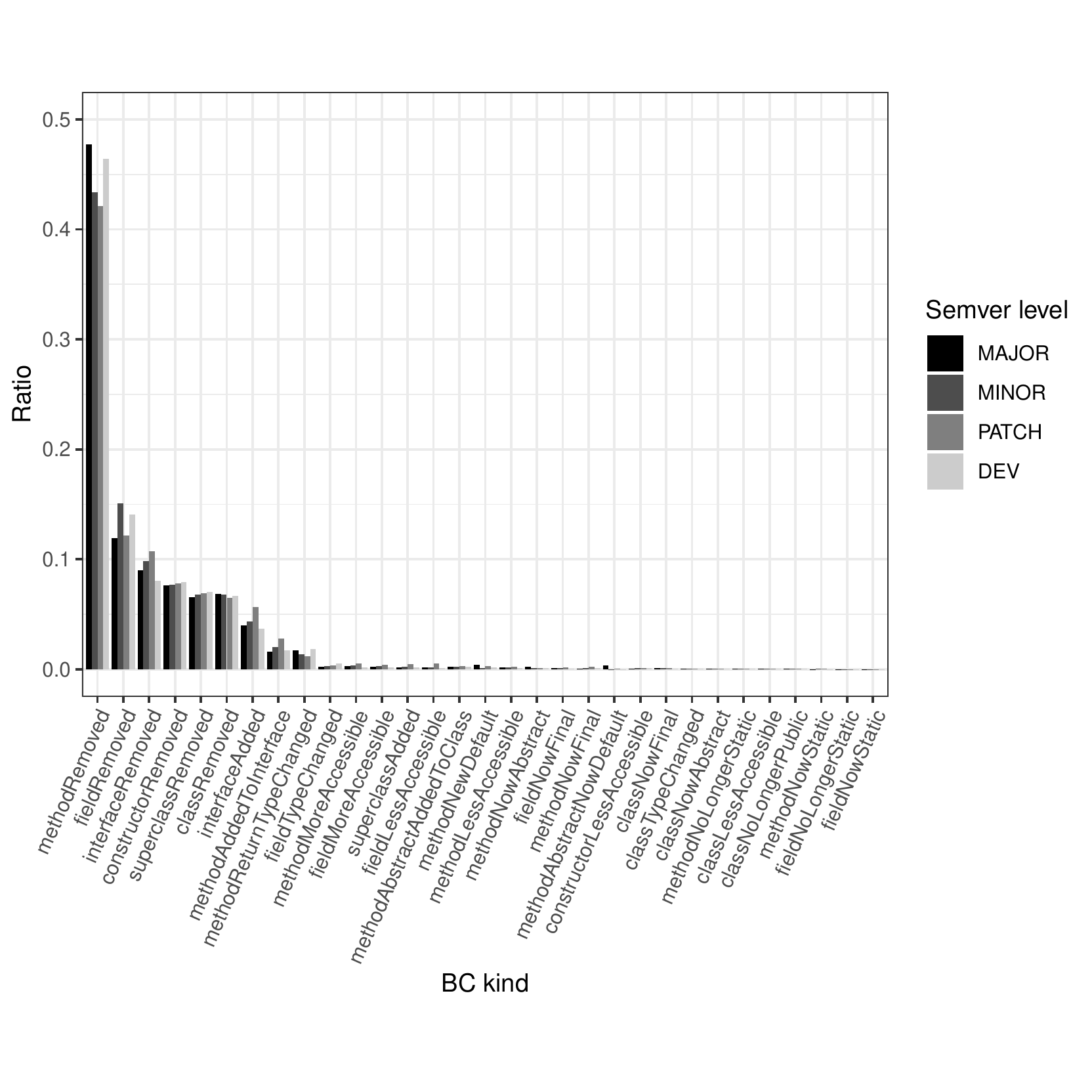}
		\caption{\dur}
	\end{subfigure}
	\caption{BC types frequency per \semver level.}
	\label{fig:bc-types}
\end{figure}

\paragraph{BC types.}
\Cref{fig:bc-types} presents the ratio of BC types (\eg \emph{method removed}, \emph{field removed}) using a bar plot, for both \duo and \dur.
BCs are discriminated by \semver levels and ordered from the most to the least frequent.
We notice that the ratio of BC types is consistent across \semver levels (except perhaps for the \emph{method return type changed} and \emph{field type changed} in the original dataset).
In both datasets, the 10 most common BC types and their associated ratios remain mostly unchanged: \emph{method removed}, \emph{field removed}, \emph{interface removed}, \emph{constructor removed}, \emph{superclass removed}, \emph{class removed}, \emph{interface added}, \emph{method added to interface}, \emph{method return type changed}, and \emph{field type changed}.
Similarly, in both datasets, the BC kinds ranked after \emph{method return type changed} are very rare.
Interestingly, the five most frequent BC kinds are all related to the removal of API entities.
We cannot directly compare these results with the \os, given that not all reported BC types are identified in the same way between the underlying tools (\ie \clirr and \japi).
However, our observations align with the results reported in the original study where method, class, and field removal headed the list.
Naturally, the BCs \emph{methodNewDefault} and \emph{methodAbstractNowDefault} do not occur in the \duo dataset, as they relate to the \ijava{default} operator which was only introduced in Java 8 (2014), while the most recent artefacts in \dur date back to 2011.

\subsubsection{Analysis}
Artefacts on \ac{MCR} do not strictly follow \semver, as an important ratio of non-major upgrades are breaking (20.1\% in \dur), confirming the main result from the original study.
However, we observe a sharp difference in the ratio of breaking upgrades per \semver level with our protocol (61.8\% of major upgrades, 37.9\% of minor upgrades, 14.6\% of patch upgrades, and 26.7\% of initial development upgrades break).
This contrasts with the original study, which reports a similar ratio of breaking upgrades for major and minor cases, with patch upgrades only slightly more stable.
In general, differences between the results reported in the \os and \duo are explained by the additional filters considered in our protocol, the increased accuracy of \mrc and \japi in detecting BCs, and the consideration of APIs annotated as unstable at the source code level (\cf~\Cref{sec:analysis-approach}).
Differences between \duo and \dur are mainly due to the increased time span and the number of artefacts.
This rationale applies to the forthcoming analyses.
Our results suggest that \semver principles are followed to some extent in practice, as 83.4\% of the library upgrades we analyse do comply with the backwards compatibility requirements of the versioning scheme.

We also notice that major upgrades not only result in a higher number of breaking cases but also tend to introduce more BCs per breaking upgrade.
Patch upgrades are the ones introducing the least number of BCs.
This suggests that, even when a non-major release is breaking, the amount of work ending on the clients' shoulders is not as high as for a major release.
Finally, the most common BCs are aligned with results presented in the original study:~ removal of API members is the most common type of BC occurring in libraries.

\begin{Result}[\rqone: How are semantic versioning principles applied in the MCR?]
	\hone asserts that ``\textit{\acp{BC} are widespread without regard for versioning principles}.'' From our analysis, we conclude that although \semver principles are not always strictly applied (20.1\% of non-major releases are breaking), they are largely followed:~83.4\% of all upgrades comply with \semver regarding backwards compatibility guarantees, and the differences between \semver levels are notable. Not only do minor and patch releases break less often than major releases, they also introduce fewer BCs. This leads us to reject \hone.
\end{Result}

\subsection{\rqtwo:~To what extent has the adherence to semantic versioning principles increased over time?}
\label{sec:rq2-protocol}

\subsubsection{Method}

To answer \rqtwo, we first study how the ratio of breaking upgrades for the various \semver levels has evolved over time, aggregated per year.
The ratio of breaking upgrades corresponds to the number of upgrades containing at least one \ac{BC} over the total number of upgrades per \semver level.
We still consider the four different \semver levels plus the analysis of non-major upgrades as a whole.
Reported results are based on the data extracted from the \duo and \dur datasets.
The latter spans fourteen years of Maven artefacts from \acs{MCR} (2005 to 2018 included).
We then contrast these results against the ones reported in the original study.
Studying the evolution of the adherence to semantic versioning principles is especially relevant as the \semver specifications are fairly recent in the history of MCR:~\semver 1.0.0 was released in 2009 and \semver 2.0.0 in 2013.
It is thus likely that the principles of \semver did not percolate yet in the dataset used in the original study.

\subsubsection{Results}

\begin{figure}
	\centering
	\includegraphics[width=1\linewidth]{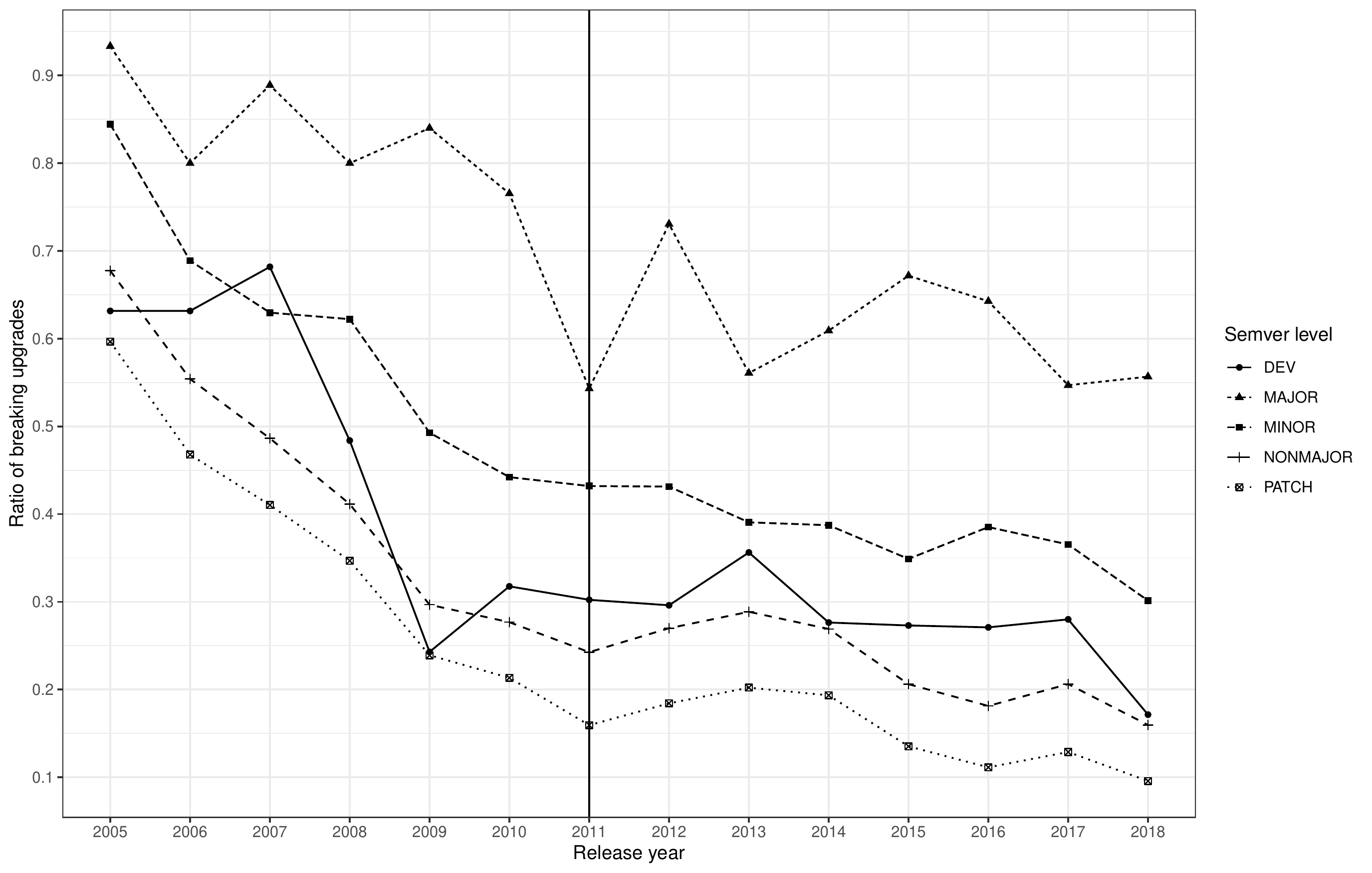}
	\caption{\rqtwo:~Evolution of the ratio of breaking upgrades per \semver level in \dur. Each data point aggregates the number of breaking upgrades of the given type for an entire year. A vertical line delimitates the periods of the original and updated datasets.}
	\label{fig:rq2-semver-year}
\end{figure}

\Cref{fig:rq2-semver-year} depicts how the ratio of breaking upgrades has evolved for major, minor, patch, initial development, and non-major levels.
Overall, the ratio of breaking upgrades, regardless of the \semver level, tends to decrease.
As expected, the ratio of breaking upgrades of major and initial development levels is more chaotic since \acp{BC} are allowed in these releases.
Nevertheless, even major and initial development releases contain fewer \acp{BC} in 2018 than in 2005.
One possible hypothesis is that clients' tolerance for \acp{BC} has decreased over time and that libraries are avoiding them more and more, even when allowed.
Additionally, different ecosystems such as npm, RubyGems, Cargo, Maven Central~\cite{decan2019what}, and GitHub are encouraging library developers to follow semver guidelines. 
In particular, GitHub explicitly states in its official documentation: ``\textit{We recommend naming tags that fit within semantic versioning.}"\footnote{\url{https://docs.github.com/en/github/administering-a-repository/releasing-projects-on-github/managing-releases-in-a-repository}}
Another relevant example is the Maven ecosystem, which offers the \texttt{maven-release-semver-policy} module to enforce the use of \semver when releasing a project.

Over 14 years, the ratio of breaking minor upgrades has decreased almost by a factor of three (from 84.4\% to 30.1\%) and the ratio of breaking patch upgrades has decreased by a factor of six (from 59.7\% to 9.6\%). 
This is to be contrasted with the results of the original study, which finds that, from 2005 to 2011, the number of non-major breaking upgrades has decreased from 28.4\% to 23.7\%.
Conversely, we find that non-major breaking upgrades have decreased from 67.7\% in 2005 to 16.0\% in 2018.

\subsubsection{Analysis}

As \mrc and \japi are able to detect more types of \acp{BC}, the percentages we report are higher than the ones reported in the original study, which makes the decrease of the ratio of breaking non-major upgrades much steeper than originally reported (a 44\% reduction instead of a 5\% decrease).
However, the decrease in the extended period is less evident: only a 9.2\% decrease.
Visually, 2011 appears as a turning point w.r.t. the decrease of breaking non-major upgrades, as the slope is less steep after this date.
Nonetheless, we found no plausible explanation for this phenomenon.
Overall, it confirms once more the statement that even though not all artefacts on \ac{MCR} follow \semver guidelines, there is an increasing tendency to comply with the versioning scheme principles.

\begin{Result}[\rqtwo: To what extent has the adherence to semantic versioning principles increased over time?]
\htwo states that ``\textit{The adherence to versioning principles has increased over time.}''
Our results confirm the results of the original study. They also show that the improvement over time is much higher than initially reported for the 2005--2011 period. The tendency persists in the 2011--2018 period, although the slope is less steep. Thus, we cannot reject \htwo.
\end{Result}

\subsection{\rqthree:~What is the impact of BCs on clients?}
\label{sec:rq4-protocol}

\subsubsection{Method}
In \rqthree, we investigate to which extent \acp{BC} introduced in Java libraries impact their clients on MCR.
More formally, for every \dmodel computed between versions $\langle v_1 \rightarrow v_2 \rangle$ of a given library (\cf \rqone), we extract all the clients $c$ declaring a compile-time or test-time dependency towards $v_1$ to uncover the impact $\Delta \langle v_1, v_2 \rangle$ would have on $c$ if it was updated to $v2$.
Concretely, we use the static analysis capabilities of \mrc to pinpoint which code locations in $c$ are impacted by individual \acp{BC} of the corresponding \dmodel (\cf \Cref{sec:mrc}).

The impact a BC has on client code varies according to if and how the client uses the declaration affected by the change (\cf \Cref{lst:bc-example}).
Hence, determining the impact of BCs requires a deep understanding of how clients and libraries interact.
For every client, we classify the impact of each individual \acs{BC} in one of three categories:~(i)~the declaration affected by the change is not used in client code (\emph{unused}); (ii)~the declaration affected by the change is used in a non-breaking way (\emph{non-breaking}), and; (iii) the declaration affected by the change is used in a breaking way (\emph{breaking}).

As with the first two research questions, we report results for both the \ac{MDD} and the \ac{MDG} corpora.
The datasets \dddo and \dddr contain, respectively, $35,539$ and $293,817$ clients which are potentially impacted by a \dmodel extracted in \rqone.
As it would be impractical to analyse these cases exhaustively, we resort to analyse a subset of them by performing a random sampling.
The question we ask for each case is: does client $c$ break when upgrading from version $v1$ to version $v2$ of a library it uses?
To answer this question with a confidence level of 99\% ($c=0.99$), an error margin of 1\% ($e=0.01$), and an estimated proportion of the population $p=0.5$ (the more conservative value yielding the largest sample size) of broken clients, we apply the standard Cochran's sample size formula to determine sample sizes for each kind of upgrade (\ie major, minor, patch, and initial development).
Then, we draw upgrades at random, without replacement, from the set of all upgrades, all major upgrades, all minor upgrades, all patch upgrades, and all development upgrades, yielding the samples depicted in \Cref{tab:dependencies-samples}.
For each tuple $\langle c, v_1, v_2 \rangle$ in the corresponding samples, we use \mrc to compute their detection models and analyse the impact of $\Delta(v_1, v_2)$ on client $c$, distinguishing among \emph{unused} declarations, \emph{non-breaking} uses, and \emph{breaking} uses.

To uncover which kinds of upgrade break clients the most, we compare the percentage of overall broken clients per \semver level.
Afterwards, we consider the number of broken locations per client for each level.

\subsubsection{Results}

\begin{table}[bt]
	\centering
	\caption{Samples derived from the population of dependencies.}
	\begin{tabular}{lrrrrr}
		& All & Major & Minor & Patch & Dev \\
		\midrule
		\multicolumn{6}{c}{\dddo} \\
		\midrule
		Population size   &  35,539 &    2,861 &  13,444 &  17,425 & 1,809   \\
		Sample size       &  11,310 &    2,440 &   7,426 &   8,498 & 1,631   \\
		Broken clients    &   1,076 &      309 &     883 &     514 &   300   \\
		\% broken clients &  9.5\% &  12.7\% & 11.9\% &  6.0\% & 18.4\% \\
		\midrule
		\multicolumn{6}{c}{\dddr} \\
		\midrule
		Population size   & 293,817 &   29,847 & 111,830 & 123,286 & 28,854  \\
		Sample size       &  15,701 &   10,663 &  14,445 &  14,621 & 10,533  \\
		Broken clients    & 1,237   &    1,250 &   1,130 &     735 &  1,772  \\
		\% broken clients & 7.9\%  & 11.7\%  &  7.8\% &  5.0\% & 16.8\% \\
	\end{tabular}
	\label{tab:dependencies-samples}
\end{table}

\paragraph{Broken clients.}
\Cref{tab:dependencies-samples} depicts the size of each \semver sample and the number and proportion of broken clients for both \dddo and \dddr. 
We observe that 9.5\% and 7.9\% of all clients for \dddo and \dddr, respectively, would break if they upgraded their dependency to the next release.

Taking into account the kind of upgrade yields interesting results:~in both datasets, initial development upgrades lead to the highest percentage of broken clients (18.4\% for \dddo and 16.8\% for \dddr), followed by major (12.7\% for \dddo and 11.7\% for \dddr), minor (11.9\% for \dddo and 7.8\% for \dddr), and finally, patch upgrades (6.0\% for \dddo and 5.0\% for \dddr).
This indicates that clients are more likely to break when upgrading to a version of a library that is potentially breaking according to \semver conventions, with initial development releases being the most problematic.
Conversely, clients that upgrade to minor and patch releases are less likely to be affected.

As we resort to random sampling to estimate the proportion of broken clients, we use statistical inference to assess our raw results.
For the sake of simplicity, we only perform the statistical analysis for the \dddr dataset.
We have the following null hypothesis: ``\textit{the proportion of broken clients is the same across each \semver level of library upgrades}''.
Note that in the remainder of this section, we use * to label the significance of the p-values using the following scale: * indicates a $p < 0.1$, ** a $p < 0.05$ and *** a $p < 0.01$.
We run a $\mathcal{X}^2$ (chi-squared) test on the table containing the amount of broken and non-broken clients for each level.
This test yields a $p < 2.2 \times 10^{-16}$ ***, therefore, we reject the null hypothesis and accept the alternative hypothesis ``\textit{the proportion of broken clients is different across each \semver level of library upgrades}''.

To assess the differences across \semver levels, we conduct post-hoc analyses for each pair of groups using Fisher's exact test on the contingency tables.
We adjust the resulting $p$-values using a Holm-Bonferroni correction.
Finally, we assess the effect size using the odds ratio.
We obtain the results shown in~\Cref{tab:fisher-results}.

\begin{table}[bt]
	\centering
	\caption{$p$-values and odds ratios across all pairs of \semver levels in \dddr.}
	\begin{tabular}{lrr}
		\semver level & $p$-value & Odds ratio\\
		\midrule
		Major vs minor   & $7.45 \times 10^{-25}$ *** & $0.64$ \\
		Major vs patch   & $3.02 \times 10^{-83}$ *** & $0.40$ \\
		Major vs dev     & $6.34 \times 10^{-26}$ *** & $1.52$ \\
		Minor vs patch   & $1.80 \times 10^{-22}$ *** & $0.62$ \\
		Minor vs dev     & $2.13 \times 10^{-104}$ *** & $2.38$ \\
		Patch vs dev     & $1.37 \times 10^{-206}$ *** & $3.82$ \\
	\end{tabular}
	\label{tab:fisher-results}
\end{table}

The $p$-values are all significant considering a $0.01$ threshold.
If we look at the direction of the odds ratios, the results are as expected w.r.t. the differences among levels.
The proportion of broken clients is higher for initial development upgrades, then major upgrades, then minor upgrades, and finally patch upgrades.
Looking at the values of the odds ratios, we note that the difference in the odds of being broken depending on the \semver level is perhaps not as high as one would expect.
For instance, for major versus minor the odds of being broken in a minor upgrade is $0.6$ times the odds of being broken in a major upgrade.
An interesting finding is that, in the Maven ecosystem, initial development upgrades break a greater proportion of clients than major upgrades.

\paragraph{Number of detections.}

\begin{figure}[t]
	\centering
	\begin{subfigure}[b]{.49\textwidth}
		\centering
		\includegraphics[width=\textwidth]{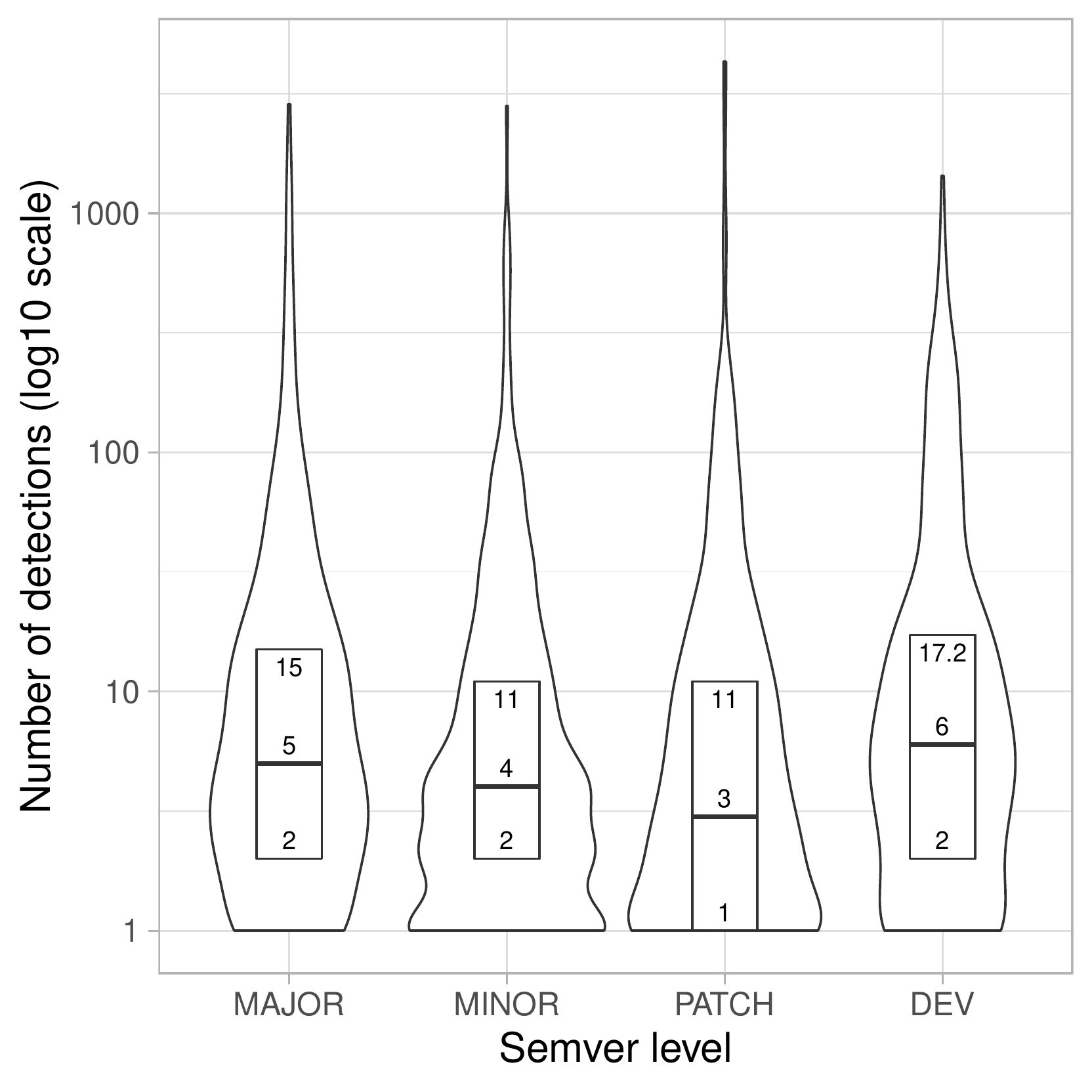}
		\caption{\dddo}
	\end{subfigure}
	\hfill
	\begin{subfigure}[b]{.49\textwidth}
		\centering
		\includegraphics[width=\textwidth]{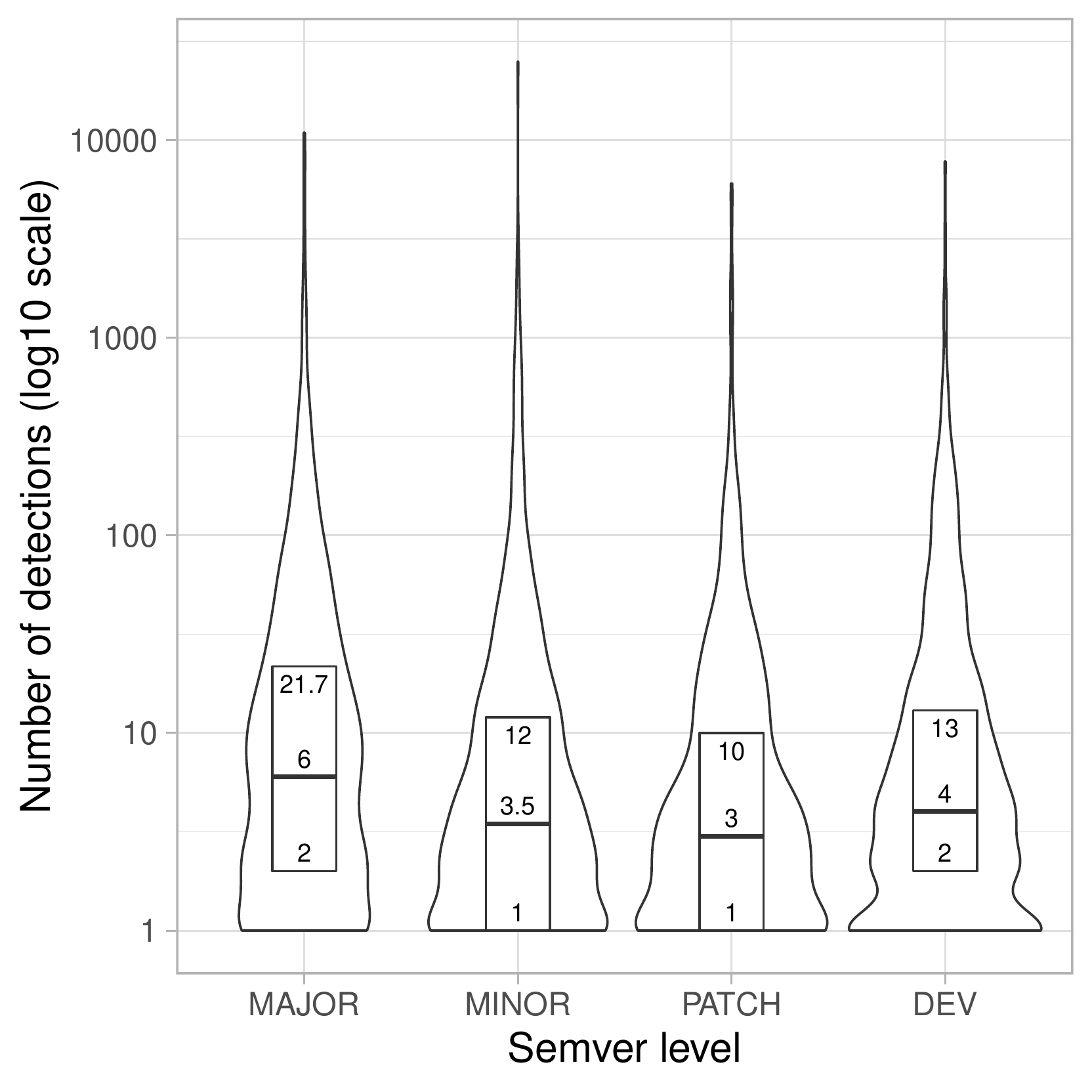}
		\caption{\dddr}
	\end{subfigure}
	\caption{Number of detections per \semver level.}
	\label{fig:detections-number}
\end{figure}

\Cref{fig:detections-number} presents the distribution in logarithmic scale of the number of broken declarations (\ie detections) per client.
Figures are presented for each \semver sample in both \dddo and \dddr.
In these distributions we only consider broken clients, that is, clients that have at least one declaration affected by a BC.
In both datasets \dddo and \dddr, we observe a similar trend:~major upgrades yield the highest number of broken declarations (medians of 5 and 6, respectively), followed by minor upgrades (medians of 4 and 3.5, respectively) and patch upgrades (medians of 3 and 3, respectively).
In \dddo, initial development upgrades yield even more broken declarations than in major upgrades (median of 6), as opposed to what we observe in \dddr (median of 4).

As we resort to random sampling to estimate the number of breaking declarations in broken clients, we use statistical inference to assess our raw results.
For the sake of simplicity, we only perform the statistical analysis for the \dddr dataset.
We have the following null hypothesis: ``\textit{the number of broken declarations is the same across each \semver level of library upgrades}''.
We run a Kruskal-Wallis rank-sum test on the number of broken declarations for each \semver level.
This test yields a $p = 3.82 \times 10^{-16}$ ***, therefore, we reject the null hypothesis and accept the alternative hypothesis ``\textit{the number of broken declarations of broken clients different across each \semver level of library upgrades}''.

To make an in-depth assessment of the differences across \semver levels, we conduct post-hoc analyses for each pair of groups, using a two-tailed Mann-Whitney test.
We adjust the resulting $p$-value using a Holm-Bonferroni correction.
In addition, we compute Cliff's delta to assess the effect size and report the interpretation of its value using Cohen's scale.
We obtain the results shown in~\Cref{tab:whitney-results}.

\begin{table}[bt]
	\centering
	\caption{$p$-values and Cliff's delta across all pairs of \semver levels in \dddr.}
	\begin{tabular}{lrr}
		\semver levels & $p$-value & Cliff's delta\\
		\midrule
		Major vs minor & $4.89 \times 10^{-11}$ *** & $0.16$ (small) \\
		Major vs patch & $2.79 \times 10^{-13}$ *** & $0.20$ (small) \\
		Major vs dev   & $2.04 \times 10^{-8}$ *** & $0.12$ (negligible)     \\
		Minor vs patch & $0.165$ & $0.04$ (negligible)  \\
		Minor vs dev   & $0.165$ & $-0.04$ (negligible) \\
		Patch vs dev   & $0.004$ *** & $-0.08$ (negligible) \\
	\end{tabular}
	\label{tab:whitney-results}
\end{table}

We note that two pairs are not significant (minor versus patch and minor versus dev), while the others are all significant at the $0.01$ threshold.
Looking at the direction of Cliff's deltas, the results are aligned with our expectations:~the number of breaking declarations in major upgrades is greater than in minor, patch, and initial development upgrades, and the number of breaking declarations in initial development upgrades is greater than in minor and patch upgrades.
Looking at the values of Cliff's deltas, however, we note that the differences are very small across the groups.
It indicates that, when a client is broken, the number of broken declarations it contains is similar whatever the \semver level of the upgrade is.

\begin{figure}
	\centering
	\includegraphics[width=\hsize]{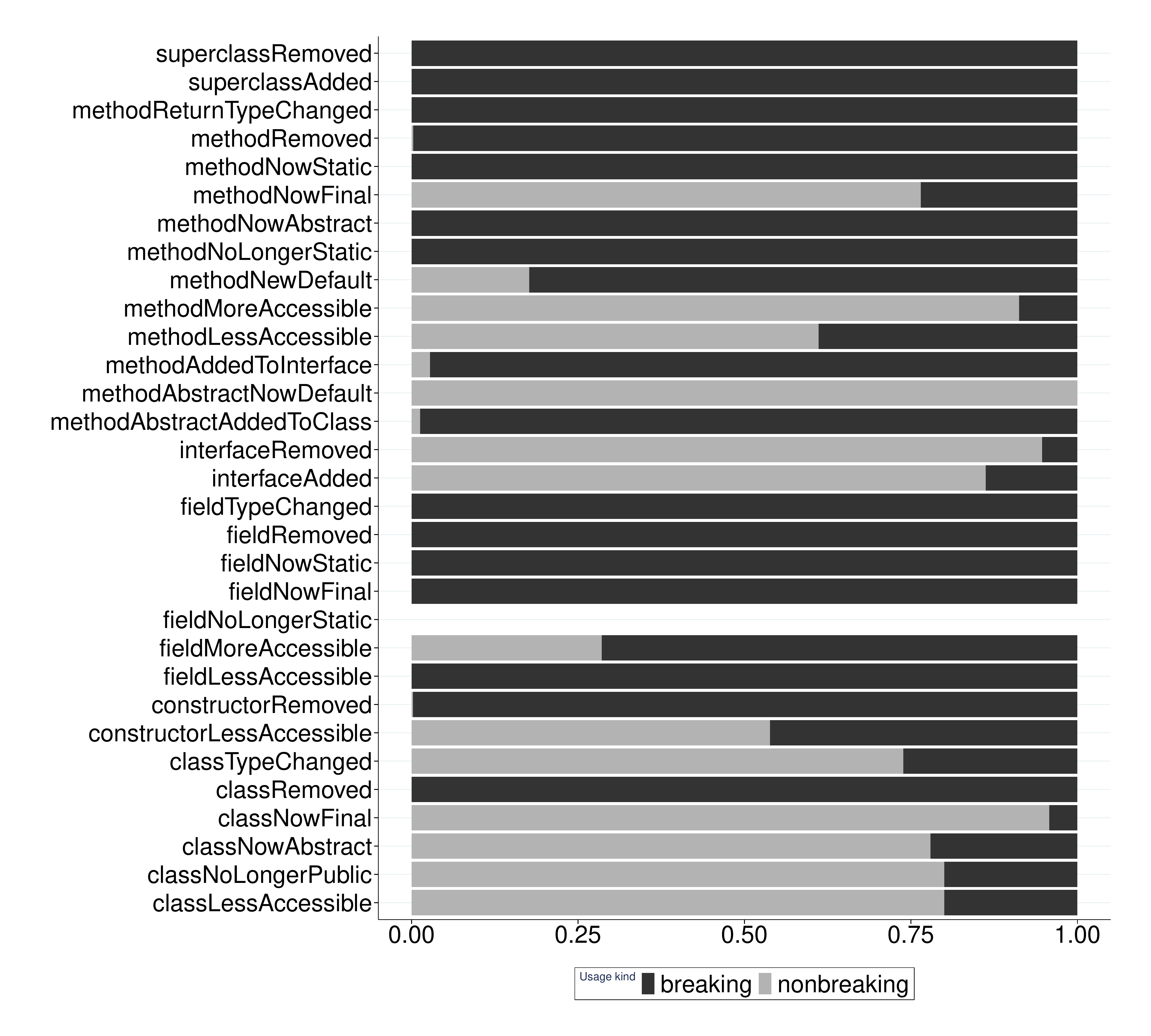}
	\caption{Ratio of breaking and non-breaking uses of API elements w.r.t. the BC type in \dddr.}
	\label{fig:mdg-bcs-uses}
\end{figure}

\paragraph{BC types.}
\Cref{fig:mdg-bcs-uses} shows the ratio of breaking and non-breaking uses of broken declarations for each BC type in \dddr.
We note that most BCs result in breaking clients as soon as they use the broken declaration.
Interestingly, we find several BCs that, in most cases, do not break clients even when the broken declaration is used in the client code.
On the other hand, apart from the \emph{interface removed} and \emph{interface added} BCs, all other popular BCs (as computed in~\Cref{sec:res-rq1}) are prone to break clients.
However, it should be noted that, for most types of BCs, there is not enough data to support a definitive conclusion.
This prevents us from proceeding to a reliable statistical analysis.

\subsubsection{Analysis}
Considering the results for the MDG corpus, we find that initial development and major releases tend to impact a higher number of clients (11.7\% and 16.8\%, respectively), as compared to minor and patch releases (7.8\% and 5.0\%, respectively).
The same tendencies can be observed in the MDD corpus.
Additionally, not only do clients break more often in major and initial development upgrades, but they also tend to break more.
In general, clients are rarely impacted by breaking declarations in the libraries they use because they do not explicitly use the affected declaration.
However, when a client uses a declaration that is affected by a \ac{BC}, it is likely to break.
These results are probably explained by the fact that library developers introduce BCs in parts of their API that are less likely to be used by their clients.
This intuition has already been investigated in the literature~\cite{harrand2019analyzing}, and our results are aligned with their observations.

\begin{Result}[\rqthree: What is the impact of breaking changes on clients?]
	\textbf{H\textsubscript{3}} asserts that ``\textit{BCs have a significant impact in terms of compilation errors in client systems}.''
	Conversely, we observe that in most cases breaking declarations are not used by client projects, which instead yields a low number of broken clients (7.9\% for all releases). The number is even lower in the case of minor and patch upgrades.
	However, when a breaking declaration is used by a client, there is a high chance that it will be impacted.
	These results contrast with those of the original study and lead us to reject \textbf{H\textsubscript{3}}.
\end{Result}

\subsection{Threats to Validity}
\label{sec:ttv}

In this section, we discuss the main threats to the validity of our replication study, following the structure recommended by \citeauthor{Wohlin2000Experimentation}~\cite{Wohlin2000Experimentation}.

\paragraph{Internal validity.}

As explained in~\Cref{sec:mrc}, the tool we implemented and used to detect \acp{BC} and their impact on client code is not perfectly accurate, which impacts the metrics we compute.
As a sanity check, we reused and extended a benchmark for Java evolution and compatibility to evaluate the accuracy of our tool, using the reference implementation of the Java linker itself as ground truth. \mrc obtained a precision of 96.3\% and a recall of 98.5\%, making the impact of this threat very low. 
Moreover, our tool is designed to be pessimistic and to over-approximate the impact of BCs in case of uncertainty.
Therefore, the impact of this threat is to slightly overestimate the number of broken clients.
\mrc does not reach 100\% recall because of two false negatives:~these are due to limitations of the underlying tool \japi which is not able to detect BCs related to the \ijava{strictfp} and \ijava{native} modifiers.
However, we do not expect that BCs related to these modifiers are common in practice.

We identified unstable API declarations of the libraries using a pre-defined list of naming conventions and annotations that was extracted by semi-automatically analysing the top-1000 most popular libraries on MCR.
However, this list is not exhaustive and cannot account for library-specific or organization-specific conventions.
As a result, we have probably misidentified some API declarations as stable. 
Assuming that unstable declarations are more likely to break than stable ones, the impact of this threat is to overestimate the number and impact of BCs.

Since it is not possible to reliably state whether a particular Maven artefact is a library or not, we consider that an artefact from MCR is a library if it has as least one \emph{external} client.
As a result, we potentially misidentified some artefacts as libraries.
Assuming that libraries are more likely to be careful about BCs than other kinds of projects, the impact of this threat is to overestimate the number and impact of BCs.

Our protocol excludes every library version suffixed with a qualifier (\eg \texttt{-beta1}, \texttt{-rc2}) as they are not final and are not meant to be used by the general public.
This complies with the Maven principle stating that every qualified version is anterior to the corresponding non-qualified version.
However, we found some libraries that always tag their versions with specific qualifiers.
For instance, the developers of the popular Google Guava library tag every version released since 2017 with a \texttt{-jre} or \texttt{-android} suffix:~there are no unqualified versions.
Google Guava versions released after 2017 are thus excluded from our datasets---even though they are legitimate, while anterior versions are included.
The impact of this threat is however low, as most of the qualifiers we found in our datasets correspond to pre-releases (\cf \Cref{sec:ds}).

The corpora we used to extract our datasets (MDD and MDG) do not contain any information regarding version ranges and constraints.
Concrete dependency versions have been resolved at the time the corpora were created, so a dependency with a version range (or no version at all) would be replaced with a concrete version picked by the Maven dependency resolver.
As a result, even if some artefacts were using version ranges, we are not able to see and analyse them.
We expect the impact of this threat to be low, as version ranges are not popular in the Maven repository.

\paragraph{External validity.}
Our study targets Java libraries and clients inside the MCR ecosystem.
Since the definition of BCs is specific to a particular programming language and since ecosystems have very different practices when it comes to \acp{BC} and versioning culture and habits~\cite{decan2019what}, there is no guarantee that our results generalize to the ecosystems of other programming languages or other Java ecosystems.

%% file: content/rw.tex
\section{Related Work}
\label{sec:rw}

Prior research in the field of library evolution has focused on understanding \textit{why} and \textit{how} evolution happens~\cite{Dig2005Role}.
Answering the \emph{why} involves understanding the motives triggering the need to change a library and its API.
In particular, researchers study the social factors motivating software change~\cite{Bavota2013Evolution, Brito2018Why, Bogart2016How, Xavier2017Why}.
Conversely, to understand \textit{how} library evolution occurs, researchers analyse the API evolution process and the evolving software itself.
In this section, we discuss a set of studies that aim at understanding \textit{how} software libraries evolve over time.
We consider studies that analyse the evolution of software ecosystems as a whole;
the nature of change in terms of backwards compatibility;
and the impact that API evolution stirs up on client projects.

\subsection{Ecosystems Evolution}
Several studies aim at understanding the evolution of a software ecosystem on its own (\eg Eclipse, Apache).
This is done to catch a glimpse of the evolution practices and expectations within the ecosystem community~\cite{Bogart2016How}.
As a direct consequence, researchers are able to create models and claims that support the development process within the studied ecosystem.

In the case of the Eclipse ecosystem, \citet{Businge2010Empirical} evaluate the applicability of Lehman's laws of software evolution to their corpus.
In a later study, the same authors analyse to which extent client plug-ins rely on unstable and internal Eclipse APIs~\cite{Businge2015Eclipse}.
They claim that more than 40\% of Eclipse plug-ins depend on this type of libraries.
Likewise, \citet{Wu2016Exploratory} study API changes and usages in both Eclipse and Apache ecosystems.
The Apache ecosystem is also studied by \citet{Bavota2013Evolution} and \citet{Raemaekers2012Measuring}. 
On the one hand, \citet{Bavota2013Evolution} report on the evolution of dependencies among projects within the ecosystem.
As the main conclusion, they discover that both the size of the ecosystem and the number of dependencies grow exponentially over time.
On the other hand, \citet{Raemaekers2012Measuring} measure Apache APIs stability with a set of metrics based on method removal, modification, and addition. 

The Squeak and Pharo ecosystems have also been a target of research.
\citet{Robbes2012How} study the ripple effects caused by method and class deprecation in these ecosystems.
They state that 14\% of the deprecated methods and 7\% of the deprecated classes impact at least one client project.
\citet{Hora2018How} complement \citet{Robbes2012How} findings.
They conclude that 61\% of client projects are impacted by API changes, and more than half of those changes trigger a reaction in affected clients.
In addition, \citet{decan2019what} perform an empirical study where they analyse the compliance to \semver principles of projects hosted in four software packaging ecosystems (\ie Cargo, npm, Packagist, and RubyGems).
They discover that certain ecosystems, such as RubyGems, do not adhere to \semver principles when analysing dependency constraints.

The abovementioned studies give a good overview of the evolution of certain ecosystems.
However, conclusions drawn by these studies do not hold outside the studied ecosystem~\cite{Sawant2016Reaction}.
Our study contributes to this body of knowledge by complementing the original results of \citet{Raemaekers2017Semantic} regarding the adherence to \semver and the impact of \acp{BC} in the Maven Central ecosystem.

\subsection{Backwards Compatibility}
The growing interest in \acp{BC} and NBCs is related to the need of analysing the stability of APIs and the impact these changes have on client projects.
One of the main observations in the literature is that backwards incompatible changes are often introduced between two versions of an API.
In fact, in a corpus of Java APIs, \citet{Dietrich2014Broken} find that 75\% of library upgrades introduce breaking changes between adjacent versions.
This study was later enhanced by \citet{Jezek2015How} who argue that 80\% of API releases are backwards incompatible.
\citet{Mostafa2017Experience} also report that 76.5\% of the releases they analyse introduce behavioural incompatible changes.
Nevertheless, there is still disagreement regarding these figures.
For instance, \citet{Xavier2017Historical} claim that only 14.78\% of the changes in their dataset are backwards incompatible, while \citet{Brito2018Why} state that 39\% of the introduced changes are classified as breaking.
These differences are due to the diversity of libraries that are analysed and their characteristics, and the criteria used to select them (for instance, the most popular Java libraries hosted on GitHub).
In this study, we detail a protocol that enables us to give a clear overview of the state of \acp{BC} in \ac{MCR} over the past 13 years.

In addition, \citet{Raemaekers2014Semantic, Raemaekers2017Semantic} conduct a study that relates \semver with backwards incompatibility.
The authors discover that \semver is not strictly followed in practice.
That is, \acp{BC} are also introduced in minor and patch releases~\cite{Jezek2015How, Raemaekers2014Semantic}.
They also claim that minor releases introduce more changes than major releases~\cite{Raemaekers2017Semantic}.
Similarly, some studies relate the nature of the API with the tendency to introduce \acp{BC}.
\citet{Xavier2017Historical} find that APIs with a higher frequency of \acp{BC} introduction tend to be more popular, larger, and active.
They also argue that the frequency of \acp{BC} increases over time.
\citet{Raemaekers2017Semantic} partially confirm this claim: larger libraries tend to introduce more \acp{BC}.
However, they also conclude that more mature APIs do not introduce more \acp{BC}, which seems counter-intuitive when contrasting the results against \citet{Xavier2017Historical} study.
\citet{decan2019what} study adherence to \semver principles at the dependency constraints level.
They find out that newer ecosystems tend to follow \semver guidelines, and that \semver practices have become more popular as time passes.
Our study complements these results by studying the adherence to \semver in Maven Central.

It is also important to be aware of the type of changes that are usually introduced during API evolution.
\citet{Cossette2012Seeking} analyse a set of binary \acp{BC} based on the affected entity type (\ie class, method, field) and its visibility (\ie protected, public).
In more recent work, \citet{Wu2016Exploratory} undergo a study that analyses 23 types of changes related to API types and methods~\cite{DesRivieres2007Evolving, Gosling2014Java}.
They find that missing classes and methods are important types of \acp{BC} affecting client projects.
\citet{Ketka2020Understanding} study type changes and required code adaptations.
They find out that type changes are more common than renamings, and that they usually appear on public entities. 
Furthermore, a particular kind of NBC has drawn much attention from the community: API deprecation~\cite{Brito2018Why, Raemaekers2014Semantic, Robbes2012How, Sawant2016Reaction}.
While deprecating an API entity does not immediately break client code, it signals that \acp{BC} may be coming in the future---as the semantics of the annotation suggests.
However, \citet{Raemaekers2014Semantic, Raemaekers2017Semantic} notice that API developers tag deprecated API entities without ever removing them from their API.
In other cases, they do quite the opposite:
API developers remove declarations from the API without deprecating them first.
\citet{Brito2018Why} point out that this is to reduce the required maintenance effort by API developers.

In spite of the contributions and findings of the abovementioned studies, there is still a long way to go.
First, some studies do not define a clear scope of the applicability of their conclusions.
In essence, it is not clear if findings account for source, binary, or behavioural incompatibility.
Moreover, how to detect and classify behavioural incompatibilities is still an open problem.
Second, the selection and study of the subset of \acp{BC} seems arbitrary, incomplete, and in some cases incorrect.
For instance, some studies concluding on backwards compatibility claim that adding a method to a class is an NBC.
Although this change is indeed binary compatible, it will break source compatibility when the method is added to an abstract class that is extended by client code.
Third, there is a lack of consensus in research findings across studies.
This is the case when reporting on the percentage of incompatible API releases and the correlation between API properties and \acp{BC} frequency.
This might be related to the underlying datasets:~some studies analyse only popular projects~\cite{Brito2018Why, Mostafa2017Experience, Xavier2017Historical}, and others consider few libraries~\cite{Kula2018Empirical, sawant2019impact}.

\subsection{Refactorings}
Refactorings are changes aimed at improving the structure of a project without changing its observable behaviour~\cite{Fowler1999Refactoring}.
One of the main inquiries concerning refactorings in API evolution is understanding to what extent API changes are actual refactorings.
\citet{Dig2005Role} discover that between 3\% and 27\% of changes on two common Java APIs are refactorings.
Furthermore, they find that at least 81\% of \acp{BC} in four Java APIs are due to refactorings.
However, in more recent studies this number might be lower.
For instance, \citet{Brito2018Why} show that 47\% confirmed \acp{BC} in their corpus are actual refactorings, and these are the most common types of \acp{BC}.
Additionally, \citet{Kula2018Empirical} state that refactorings break less than 37\% of all clients of a given API.
They find a tendency to find more \acp{BC} and refactorings in API internal entities~\cite{Brito2018Why, Kula2018Empirical}.

The main limitation of these studies is the level of abstraction at which the analysis is performed.
That is, a refactoring might be composed of multiple \acp{BC} and, in some cases, by multiple refactorings.
For instance, method renamed could be recorded in a delta as both a method removed and method added change.
Analysing these changes requires an additional effort in dissecting compound cases.

\subsection{Impact of API Evolution}
More recent studies attempt to understand how client projects are impacted by API evolution.
In some of them~\cite{Bavota2013Evolution, Dietrich2014Broken, Robbes2012How, Xavier2017Historical} we find the same claim: 
there is no massive impact of API changes on client code.
In fact, \citet{Bavota2013Evolution} state that only around 5\% of the projects in their corpus are impacted by API evolution;
\citet{Xavier2017Historical} discover that only 2.54\% client projects in the dataset are impacted by API \acp{BC}; and
\citet{Robbes2012How} show that 14\% and 7\% of class and method deprecations, respectively, impact client projects.
From a different perspective, \citet{Kula2018Empirical} state that \acp{BC} are more likely to appear in API entities that are not used by client code.
Later, \citet{Raemaekers2017Semantic} relate the number of compilation errors with the introduced \acp{BC}. 
They do so by individually inserting \acp{BC} in the API source code.

Regarding API-client co-evolution, there is a growing interest in understanding \textit{why}, \textit{when}, and \textit{how} client projects upgrade to a newer version of an API.
On the quest of answering the \textit{why} and \textit{when}, \citet{Raemaekers2014Semantic} show that API upgrading tends to be performed when major API updates are released.
\citet{Bavota2013Evolution} indirectly confirm this claim by stating that client projects upgrade to a newer version of an API only when substantial changes are introduced.
Regarding the \textit{how}, \citet{Bavota2013Evolution} highlight that even though few client projects might be affected by API evolution, certain dependencies that offer cross-cutting services can strongly impact them.
To support these first insights, \citet{Robbes2012How} find that resolving the first 25\% ripple effects in the Squeak and Pharo ecosystem, requires more than 14 developers.
In addition, several commits are registered to resolve the issue, which suggests the existence of non-trivial changes.
Both \citet{Robbes2012How} and \citet{Sawant2016Reaction} claim that finding systematic changes in affected client code is rare.
There are many cases where impacted code is simply dropped, or an ad-hoc solution is provided.
These findings are contrary to what developers postulate in \citet{Brito2018Why} study:
they argue that API migration results in minor and easy changes.
In addition, \citet{Mostafa2017Experience} claim that 67\% of bugs introduced by behavioural changes can be fixed by simple changes (\eg replacing arguments, converting return values).

Despite the new contributions in the field, few papers study both how APIs evolve and how this evolution impacts client projects.
Moreover, when they analyse API usage there is a misalignment between the API change and usage types.
For instance, \citet{Wu2016Exploratory} label \textit{inheritance for \ac{IoC}} as an API usage type.
However, this type of usage can be split into other categories, such as \textit{method overriding}, \textit{class extension}, and \textit{interface implementation}.
With this differentiation it is possible to relate API changes at different levels (\ie class, method, and field levels) with atomic API usages; and accurately point to affected client members. 
Finally, as in the case of studies related to backwards compatibility, we perceive contradicting results that are most likely due to differences in the studied datasets.

%% file: content/discussion.tex
\section{Discussion}
\label{sec:discussion}
In this section, we discuss the main implications of our findings for library developers, library clients, and researchers.

\paragraph{Implications for library developers.}
The introduction of BCs is inherent to software evolution and cannot always be avoided.
Although libraries are encouraged to preserve backwards compatibility, the need to introduce new features and improve the quality of the library sometimes results in incompatible changes.
We claim that introducing BCs is tolerable as long as they are properly announced in advance to not take clients by surprise.
Versioning conventions or code-level mechanisms (\eg annotations, naming conventions) are regularly used for this purpose.
In addition, many software ecosystems such as Maven, GitHub, and npm encourage their users to adhere to the \semver policies.
They even offer core tooling to support developers on this quest for quality and compliance.
The semantics of such policies might differ among ecosystems, thus, library developers are exhorted to carefully understand and follow such rules.

Introducing too many BCs may, however, hurt the reputation of a library.
It is very hard for developers to manually detect BCs in source code, so we believe that the use of tools such as \japi and \mrc is important to help library maintainers make the right decisions.
The Apache Commons developers, for instance, use \japi on every new release of their libraries to check for backwards compatibility.\footnote{\url{https://garygregory.wordpress.com/2020/06/14/how-we-handle-binary-compatibility-at-apache-commons}}
These tools are also able to automatically generate reports that help clients to anticipate the changes.\footnote{\url{https://commons.apache.org/proper/commons-lang/japicmp}}
The first step towards a more disciplined evolution of libraries is to detect and communicate on BCs, two aspects addressed by these tools.
As we have shown in our analysis of \rqthree, some BCs appear to be more critical than others for client developers.
We thus encourage library developers to interpret the output of these tools wisely and to account for the severity of different changes.

The ability of \mrc to infer the impact of BCs on client code is also beneficial for library developers.
Developers may run \mrc as part of a continuous integration pipeline to check that a particular commit, pull request, or release does not significantly impact their clients, and reconsider the changes if the impact is too high.
This type of tooling can also yield valuable information on how unstable or unsafe parts of APIs are being used by client projects.
With this information, library developers can either decide to promote internal interfaces to public ones, including new features in their list of public interfaces~\cite{hora2016when}.
They can also analyse to what degree unsafe declarations are impacting client projects, and based on these results, they can even come up with new designs to transform an unsafe interface into a safe one~\cite{mastrangelo2015use}.

\paragraph{Implications for client developers.}
The main implication of our results for client developers is that the situation is not as bad as reported in the literature.
In MCR, most releases comply with \semver requirements and avoid BCs in non-major releases.
Besides, as we have shown in \rqtwo, the situation has significantly improved over time.

Each ecosystem has its own policy regarding versioning conventions and the treatment of BCs.
Cargo, npm, Packagist, and Rubygems, for instance, do not apply semantic versioning in the same way~\cite{decan2019what}.
Client developers should thus pay attention to ecosystem-specific guidelines and pick an ecosystem that advocates a strict policy to minimize the risk of being impacted by unwanted changes.
Additionally, identifying unstable declarations in used APIs via \semver or other code-level mechanisms is important to avoid client broken code after upgrading to more recent releases.
Naming conventions~\cite{businge2019stable} and use of annotations are some of the signaling mechanisms that client developers should look for.

Tools such as \mrc should also be beneficial to client developers.
When faced with the possibility of upgrading a dependency, developers may employ \mrc to evaluate the impact of different versions, and choose the one that addresses their requirements without causing too much disruption.

\paragraph{Implications for researchers.}
As we have seen, the use of code-level mechanisms to delimit unstable APIs relies on conventions that might vary from one organization to the other or from one library to the other (\texttt{@Beta} for Google, \texttt{@Internal} for Apache, \texttt{sun.*} packages in the JDK, \etc).
Contrary to semantic versioning, there is no standardization of these mechanisms, which makes it hard for clients to understand which declarations should be considered stable or unstable.
Besides, it is not clear how \semver and code-level mechanisms interact:~the \semver specification only mentions the \texttt{@Deprecated} annotation.
Should developers release a major revision when they introduce a BC in a beta-stage API?
We believe that clarifying the role of code-level mechanisms and their relation with \semver would be beneficial.
Another interesting line of work would be to incorporate better mechanisms to delimit APIs directly in programming languages:~developers are currently forced to make some declarations public only for technical reasons (\cf \Cref{sec:background}) even though they are not part of the intended API.

Researchers should also strive to design and implement benchmarks to compare tools related to library evolution objectively.
The benchmark of \citeauthor{jezek2017api}~\cite{jezek2017api}, which we reuse and extend to evaluate the accuracy of \mrc, is the first step in this direction and should be complemented with other benchmarks, for instance, related to behavioural compatibility.

Furthermore, when analysing software evolution, the design of a study protocol and the creation of the underlying datasets should be carefully performed.
Sampling bias is a recurrent threat to validity that can hurt the interpretation of API evolution studies.
For instance, selecting only the most popular libraries on a repository, or only the ones related to a particular ecosystem hurts the generalization of the study findings.
To cope with this issue, representative and diverse samples are required to come up with relevant conclusions~\cite{Nagappan2013Diversity}.

Finally, because BCs are not always avoidable, researchers should continue to develop tools and methods that assist client developers in automatically migrating their code~\cite{Cossette2012Seeking, Xi2019Migrating, Xu2019Meditor}.

%% file: content/conclusion.tex
\section{Conclusion}
\label{sec:conclu}

In this paper, we conduct an external and differentiated replication study of the work presented by \citet{Raemaekers2017Semantic}.
The motivation behind this study is to better understand which kind of BCs happen in libraries hosted on the \ac{MCR}, and what is their impact.
We rely on \semver principles to draw conclusions that are aligned with versioning conventions that signal API instability.
Our protocol addresses some limitations of the original study and expands the analysis to a new dataset spanning seven more years of the \ac{MCR}.
We implement and use \mrc to compute BCs between adjacent versions of libraries, and to detect locations in client code that are affected by such BCs.

The main results of the study are as follows:
\begin{description}
	\item[\rqone:] \textbf{How are semantic versioning principles applied in the MCR?} 83.4\% of all upgrades on MCR do comply with \semver principles. Still, 20.1\% of non-major releases are breaking, threatening client projects.
	
	\item[\rqtwo:] \textbf{To what extent has the adherence to semantic versioning principles increased over time?} The tendency to comply with \semver practices has significantly increased over time:~the number of non-major breaking releases has decreased from 67.7\% in 2005 to 16.0\% in 2018.
	
	\item[\rqthree:] \textbf{What is the impact of breaking changes on clients?} Only 7.9\% of the clients we analyse are impacted by the BCs introduced in adjacent library releases. However, when breaking declarations are used by client projects, they are likely to break. 
\end{description}

According to these results, we state that libraries and client projects on the Maven ecosystem are \emph{not} ``breaking bad''. 
To be precise, developers of Maven projects tend to follow \semver principles and are for the most part disciplined when introducing BCs.
While the situation has improved over time, there is still room for improvement. 
Although the impact of BCs on client projects is low, more research is needed to support clients that are impacted and need to migrate their code.
Differences with results reported in the original study are explained by major changes introduced in the protocol and the extended time span of the new corpus.

As future work, we first would like to perform qualitative analyses to complement our findings.
In particular, we would like to explain the phenomenon we observed:~what are the motivations behind inserting \acp{BC} in non-major releases, and why has the adherence to semantic versioning increased so significantly.
These questions could be answered by interviewing library maintainers and clients.
Second, we would like to study how the evolution of the Java language itself impacts the definition of \acp{BC}, and how this affects libraries and clients.
As new constructs are made available in Java (\eg the \ijava{default} operator in Java 8 or the \ijava{record} data type in Java 15), new BCs appear.
At the same time, these new constructs provide new strategies to deal with certain BCs (\eg \ijava{default} methods allow to gracefully evolve an interface without forcing changes in existing implementations).
Third, we believe that the understanding of how client projects react to \acp{BC} is another step towards finding a way to support library-client co-evolution. 
Thus, we aspire to study how clients react in the wild and which patterns can be identified from these reactions.
Finally, we also would like to study behavioral incompatible changes in Java libraries.

%% file: main.bbl
\begin{thebibliography}{52}
\providecommand{\natexlab}[1]{#1}
\providecommand{\url}[1]{{#1}}
\providecommand{\urlprefix}{URL }
\expandafter\ifx\csname urlstyle\endcsname\relax
  \providecommand{\doi}[1]{DOI~\discretionary{}{}{}#1}\else
  \providecommand{\doi}{DOI~\discretionary{}{}{}\begingroup
  \urlstyle{rm}\Url}\fi
\providecommand{\eprint}[2][]{\url{#2}}

\bibitem[{Basten et~al.(2015)Basten, Hills, Klint, Landman, Shahi, Steindorfer,
  and Vinju}]{Basten2015M3}
Basten B, Hills M, Klint P, Landman D, Shahi A, Steindorfer MJ, Vinju JJ (2015)
  {M3:} {A} general model for code analytics in rascal. In: 1st {IEEE}
  International Workshop on Software Analytics, {SWAN} 2015, Montreal, QC,
  Canada, March 2, 2015, {IEEE} Computer Society, pp 25--28,
  \doi{10.1109/SWAN.2015.7070485}

\bibitem[{Bavota et~al.(2013)Bavota, Canfora, Penta, Oliveto, and
  Panichella}]{Bavota2013Evolution}
Bavota G, Canfora G, Penta MD, Oliveto R, Panichella S (2013) The evolution of
  project inter-dependencies in a software ecosystem: The case of apache. In:
  2013 {IEEE} International Conference on Software Maintenance, Eindhoven, The
  Netherlands, September 22-28, 2013, {IEEE} Computer Society, pp 280--289,
  \doi{10.1109/ICSM.2013.39}

\bibitem[{Benelallam et~al.(2018)Benelallam, Harrand, Valero, Baudry, and
  Barais}]{benelallam2018mdg}
Benelallam A, Harrand N, Valero CS, Baudry B, Barais O (2018) {Maven Dentral
  Dependency Graph}. \doi{10.5281/zenodo.1489120}

\bibitem[{Benelallam et~al.(2019)Benelallam, Harrand, Soto{-}Valero, Baudry,
  and Barais}]{benelallam2019maven}
Benelallam A, Harrand N, Soto{-}Valero C, Baudry B, Barais O (2019) The maven
  dependency graph: a temporal graph-based representation of maven central. In:
  Proceedings of the 16th International Conference on Mining Software
  Repositories, {MSR} 2019, 26-27 May 2019, Montreal, Canada, {IEEE} / {ACM},
  pp 344--348, \doi{10.1109/MSR.2019.00060}

\bibitem[{Bogart et~al.(2016)Bogart, K{\"{a}}stner, Herbsleb, and
  Thung}]{Bogart2016How}
Bogart C, K{\"{a}}stner C, Herbsleb JD, Thung F (2016) How to break an {API:}
  cost negotiation and community values in three software ecosystems. In:
  Proceedings of the 24th {ACM} {SIGSOFT} International Symposium on
  Foundations of Software Engineering, {FSE} 2016, Seattle, WA, USA, November
  13-18, 2016, {ACM}, pp 109--120, \doi{10.1145/2950290.2950325}

\bibitem[{Brito et~al.(2018)Brito, Xavier, Hora, and Valente}]{Brito2018Why}
Brito A, Xavier L, Hora AC, Valente MT (2018) Why and how java developers break
  apis. In: 25th International Conference on Software Analysis, Evolution and
  Reengineering, {SANER} 2018, Campobasso, Italy, March 20-23, 2018, {IEEE}
  Computer Society, pp 255--265, \doi{10.1109/SANER.2018.8330214}

\bibitem[{Businge et~al.(2010)Businge, Serebrenik, and van~den
  Brand}]{Businge2010Empirical}
Businge J, Serebrenik A, van~den Brand M (2010) An empirical study of the
  evolution of eclipse third-party plug-ins. In: Proceedings of the Joint
  {ERCIM} Workshop on Software Evolution {(EVOL)} and International Workshop on
  Principles of Software Evolution (IWPSE), Antwerp, Belgium, September 20-21,
  2010, {ACM}, pp 63--72, \doi{10.1145/1862372.1862389}

\bibitem[{Businge et~al.(2015)Businge, Serebrenik, and van~den
  Brand}]{Businge2015Eclipse}
Businge J, Serebrenik A, van~den Brand MGJ (2015) Eclipse {API} usage: the good
  and the bad. Softw Qual J 23:107--141, \doi{10.1007/s11219-013-9221-3}

\bibitem[{Businge et~al.(2019)Businge, Kawuma, Openja, Bainomugisha, and
  Serebrenik}]{businge2019stable}
Businge J, Kawuma S, Openja M, Bainomugisha E, Serebrenik A (2019) How stable
  are eclipse application framework internal interfaces? In: 26th {IEEE}
  International Conference on Software Analysis, Evolution and Reengineering,
  {SANER} 2019, Hangzhou, China, February 24-27, 2019, {IEEE}, pp 117--127,
  \doi{10.1109/SANER.2019.8668018}

\bibitem[{Cossette and Walker(2012)}]{Cossette2012Seeking}
Cossette BE, Walker RJ (2012) {Seeking the Ground Truth: A Retroactive Study on
  the Evolution and Migration of Software Libraries}. In: 20th International
  Symposium on the Foundations of Software Engineering, ACM, New York, pp
  55:1--55:11, \doi{10.1145/2393596.2393661}

\bibitem[{Decan and Mens(2021)}]{decan2019what}
Decan A, Mens T (2021) What do package dependencies tell us about semantic
  versioning? {IEEE} Trans Software Eng 47:1226--1240,
  \doi{10.1109/TSE.2019.2918315}

\bibitem[{Des~Rivi\`{e}res(2007)}]{DesRivieres2007Evolving}
Des~Rivi\`{e}res J (2007) {Evolving Java-based APIs}.
  \url{https://tinyurl.com/yyqguo34}, {last access 26.07.2019}

\bibitem[{Dietrich et~al.(2014)Dietrich, Jezek, and Brada}]{Dietrich2014Broken}
Dietrich J, Jezek K, Brada P (2014) Broken promises: An empirical study into
  evolution problems in java programs caused by library upgrades. In: 2014
  Software Evolution Week - {IEEE} Conference on Software Maintenance,
  Reengineering, and Reverse Engineering, {CSMR-WCRE} 2014, Antwerp, Belgium,
  February 3-6, 2014, {IEEE} Computer Society, pp 64--73,
  \doi{10.1109/CSMR-WCRE.2014.6747226}

\bibitem[{Dig and Johnson(2006)}]{Dig2006How}
Dig D, Johnson R (2006) {How Do APIs Evolve? A Story of Refactoring: Research
  Articles}. J Softw Maint Evol 18(2):83--107, \doi{10.1002/smr.v18:2}

\bibitem[{Dig and Johnson(2005)}]{Dig2005Role}
Dig D, Johnson RE (2005) The role of refactorings in {API} evolution. In: 21st
  {IEEE} International Conference on Software Maintenance {(ICSM} 2005), 25-30
  September 2005, Budapest, Hungary, {IEEE} Computer Society, pp 389--398,
  \doi{10.1109/ICSM.2005.90}

\bibitem[{Fowler(1999)}]{Fowler1999Refactoring}
Fowler M (1999) {Refactoring: Improving the Design of Existing Code}.
  Addison-Wesley Longman Publishing Co., Inc., Boston

\bibitem[{Godfrey and Germ{\'{a}}n(2014)}]{Godfrey2014Evolution}
Godfrey MW, Germ{\'{a}}n DM (2014) On the evolution of lehman's laws. J Softw
  Evol Process 26:613--619, \doi{10.1002/smr.1636}

\bibitem[{Gonz{\'{a}}lez{-}Barahona et~al.(2017)Gonz{\'{a}}lez{-}Barahona,
  Sherwood, Robles, and Izquierdo{-}Cortazar}]{gonzalez2017technical}
Gonz{\'{a}}lez{-}Barahona JM, Sherwood P, Robles G, Izquierdo{-}Cortazar D
  (2017) Technical lag in software compilations: Measuring how outdated a
  software deployment is. In: Open Source Systems: Towards Robust Practices -
  13th {IFIP} {WG} 2.13 International Conference, {OSS} 2017, Buenos Aires,
  Argentina, May 22-23, 2017, Proceedings, vol 496, pp 182--192,
  \doi{10.1007/978-3-319-57735-7\\_17}

\bibitem[{Gosling et~al.(2014)Gosling, Joy, Steele, Bracha, and
  Buckley}]{Gosling2014Java}
Gosling J, Joy B, Steele GL, Bracha G, Buckley A (2014) {The Java Language
  Specification, Java SE 8 Edition}, 1st edn. Addison-Wesley Professional

\bibitem[{Harrand et~al.(2019)Harrand, Benelallam, Soto{-}Valero, Barais, and
  Baudry}]{harrand2019analyzing}
Harrand N, Benelallam A, Soto{-}Valero C, Barais O, Baudry B (2019) Analyzing
  2.3 million maven dependencies to reveal an essential core in apis. arXiv
  preprint arXiv:190809757 abs/1908.09757

\bibitem[{Hora et~al.(2016)Hora, Valente, Robbes, and Anquetil}]{hora2016when}
Hora A, Valente MT, Robbes R, Anquetil N (2016) {When Should Internal
  Interfaces Be Promoted to Public?} In: 2016 24th ACM SIGSOFT International
  Symposium on Foundations of Software Engineering, ACM, New York, p 278–289,
  \doi{10.1145/2950290.2950306}

\bibitem[{Hora et~al.(2018)Hora, Robbes, Valente, Anquetil, Etien, and
  Ducasse}]{Hora2018How}
Hora AC, Robbes R, Valente MT, Anquetil N, Etien A, Ducasse S (2018) How do
  developers react to {API} evolution? {A} large-scale empirical study. Softw
  Qual J 26:161--191, \doi{10.1007/s11219-016-9344-4}

\bibitem[{Jezek and Dietrich(2017)}]{jezek2017api}
Jezek K, Dietrich J (2017) {API} evolution and compatibility: {A} data corpus
  and tool evaluation. J Object Technol 16:2:1--23,
  \doi{10.5381/jot.2017.16.4.a2}

\bibitem[{Jezek et~al.(2015)Jezek, Dietrich, and Brada}]{Jezek2015How}
Jezek K, Dietrich J, Brada P (2015) How java apis break - an empirical study.
  Inf Softw Technol 65:129--146, \doi{10.1016/j.infsof.2015.02.014}

\bibitem[{Ketkar et~al.(2020)Ketkar, Tsantalis, and
  Dig}]{Ketka2020Understanding}
Ketkar A, Tsantalis N, Dig D (2020) Understanding type changes in java. In:
  {ESEC/FSE} '20: 28th {ACM} Joint European Software Engineering Conference and
  Symposium on the Foundations of Software Engineering, Virtual Event, USA,
  November 8-13, 2020, {ACM}, pp 629--641, \doi{10.1145/3368089.3409725}

\bibitem[{Klint et~al.(2010)Klint, van~der Storm, and Vinju}]{rascal}
Klint P, van~der Storm T, Vinju JJ (2010) {EASY Meta-Programming with Rascal.
  Leveraging the Extract-Analyze-SYnthesize Paradigm for Meta-Programming}. In:
  3rd International Summer School on Generative and Transformational Techniques
  in Software Engineering, Springer, LNCS

\bibitem[{Kula et~al.(2018{\natexlab{a}})Kula, Germ{\'{a}}n, Ouni, Ishio, and
  Inoue}]{gaikovina2017developers}
Kula RG, Germ{\'{a}}n DM, Ouni A, Ishio T, Inoue K (2018{\natexlab{a}}) Do
  developers update their library dependencies? - an empirical study on the
  impact of security advisories on library migration. Empir Softw Eng
  23:384--417, \doi{10.1007/s10664-017-9521-5}

\bibitem[{Kula et~al.(2018{\natexlab{b}})Kula, Ouni, Germ{\'{a}}n, and
  Inoue}]{Kula2018Empirical}
Kula RG, Ouni A, Germ{\'{a}}n DM, Inoue K (2018{\natexlab{b}}) An empirical
  study on the impact of refactoring activities on evolving client-used apis.
  Inf Softw Technol 93:186--199, \doi{10.1016/j.infsof.2017.09.007}

\bibitem[{Kühne et~al.(2003)Kühne, Massol, and Kitching}]{Kuhne2003Clirr}
Kühne L, Massol V, Kitching S (2003) {The Clirr Maven Plugin}.
  \url{https://tinyurl.com/y6az94l4}, {last access 08.04.2020}

\bibitem[{Lehman(1978)}]{Lehman1978Programs}
Lehman MM (1978) {Programs, Cities, Students--- Limits to Growth?}, Springer,
  New York, pp 42--69. \doi{10.1007/978-1-4612-6315-9\_6}

\bibitem[{Lehman et~al.(1997)Lehman, Ramil, Wernick, Perry, and
  Turski}]{Lehman1997Metrics}
Lehman MM, Ramil JF, Wernick P, Perry DE, Turski WM (1997) Metrics and laws of
  software evolution - the nineties view. In: 4th {IEEE} International Software
  Metrics Symposium {(METRICS} 1997), November 5-7, 1997, Albuquerque, NM,
  {USA}, {IEEE} Computer Society, p~20, \doi{10.1109/METRIC.1997.637156}

\bibitem[{Lindsay~Murray and Ehrenberg(1993)}]{Lindsay1993Design}
Lindsay~Murray R, Ehrenberg ASC (1993) {The Design of Replicated Studies}. The
  American Statistician 47(3):217--228, \doi{10.2307/2684982}

\bibitem[{Mastrangelo et~al.(2015)Mastrangelo, Ponzanelli, Mocci, Lanza,
  Hauswirth, and Nystrom}]{mastrangelo2015use}
Mastrangelo L, Ponzanelli L, Mocci A, Lanza M, Hauswirth M, Nystrom N (2015)
  Use at your own risk: the java unsafe {API} in the wild. In: Proceedings of
  the 2015 {ACM} {SIGPLAN} International Conference on Object-Oriented
  Programming, Systems, Languages, and Applications, {OOPSLA} 2015, part of
  {SPLASH} 2015, Pittsburgh, PA, USA, October 25-30, 2015, {ACM}, pp 695--710,
  \doi{10.1145/2814270.2814313}

\bibitem[{Mirhosseini and Parnin(2017)}]{mirhosseini2017automated}
Mirhosseini S, Parnin C (2017) Can automated pull requests encourage software
  developers to upgrade out-of-date dependencies? In: Proceedings of the 32nd
  {IEEE/ACM} International Conference on Automated Software Engineering, {ASE}
  2017, Urbana, IL, USA, October 30 - November 03, 2017, {IEEE} Computer
  Society, pp 84--94, \doi{10.1109/ASE.2017.8115621}

\bibitem[{Mostafa et~al.(2017)Mostafa, Rodriguez, and
  Wang}]{Mostafa2017Experience}
Mostafa S, Rodriguez R, Wang X (2017) Experience paper: a study on behavioral
  backward incompatibilities of java software libraries. In: Proceedings of the
  26th {ACM} {SIGSOFT} International Symposium on Software Testing and
  Analysis, Santa Barbara, CA, USA, July 10 - 14, 2017, {ACM}, pp 215--225,
  \doi{10.1145/3092703.3092721}

\bibitem[{Nagappan et~al.(2013)Nagappan, Zimmermann, and
  Bird}]{Nagappan2013Diversity}
Nagappan M, Zimmermann T, Bird C (2013) Diversity in software engineering
  research. In: Joint Meeting of the European Software Engineering Conference
  and the {ACM} {SIGSOFT} Symposium on the Foundations of Software Engineering,
  ESEC/FSE'13, Saint Petersburg, Russian Federation, August 18-26, 2013, {ACM},
  pp 466--476, \doi{10.1145/2491411.2491415}

\bibitem[{Preston-Werner(2013)}]{Preston2013Semver}
Preston-Werner T (2013) {Semantic Versioning 2.0.0}.
  \url{https://tinyurl.com/y7t7g6t5}, {last access 30.07.2019}

\bibitem[{Raemaekers(2013)}]{Raemaekers2013Dataset}
Raemaekers S (2013) {The Maven Dependency Dataset}.
  \url{https://tinyurl.com/uveepue}, {last access 09.04.2020}

\bibitem[{Raemaekers et~al.(2012)Raemaekers, van Deursen, and
  Visser}]{Raemaekers2012Measuring}
Raemaekers S, van Deursen A, Visser J (2012) Measuring software library
  stability through historical version analysis. In: 28th {IEEE} International
  Conference on Software Maintenance, {ICSM} 2012, Trento, Italy, September
  23-28, 2012, {IEEE} Computer Society, pp 378--387,
  \doi{10.1109/ICSM.2012.6405296}

\bibitem[{Raemaekers et~al.(2013)Raemaekers, van Deursen, and
  Visser}]{Raemaekers2013Maven}
Raemaekers S, van Deursen A, Visser J (2013) The maven repository dataset of
  metrics, changes, and dependencies. In: Proceedings of the 10th Working
  Conference on Mining Software Repositories, {MSR} '13, San Francisco, CA,
  USA, May 18-19, 2013, {IEEE} Computer Society, pp 221--224,
  \doi{10.1109/MSR.2013.6624031}

\bibitem[{Raemaekers et~al.(2014)Raemaekers, van Deursen, and
  Visser}]{Raemaekers2014Semantic}
Raemaekers S, van Deursen A, Visser J (2014) Semantic versioning versus
  breaking changes: {A} study of the maven repository. In: 14th {IEEE}
  International Working Conference on Source Code Analysis and Manipulation,
  {SCAM} 2014, Victoria, BC, Canada, September 28-29, 2014, {IEEE} Computer
  Society, pp 215--224, \doi{10.1109/SCAM.2014.30}

\bibitem[{Raemaekers et~al.(2017)Raemaekers, van Deursen, and
  Visser}]{Raemaekers2017Semantic}
Raemaekers S, van Deursen A, Visser J (2017) Semantic versioning and impact of
  breaking changes in the maven repository. J Syst Softw 129:140--158,
  \doi{10.1016/j.jss.2016.04.008}

\bibitem[{Robbes et~al.(2012)Robbes, Lungu, and
  R{\"{o}}thlisberger}]{Robbes2012How}
Robbes R, Lungu M, R{\"{o}}thlisberger D (2012) How do developers react to
  {API} deprecation?: the case of a smalltalk ecosystem. In: 20th {ACM}
  {SIGSOFT} Symposium on the Foundations of Software Engineering (FSE-20),
  SIGSOFT/FSE'12, Cary, NC, {USA} - November 11 - 16, 2012, {ACM}, p~56,
  \doi{10.1145/2393596.2393662}

\bibitem[{Sawant(2019)}]{sawant2019impact}
Sawant AA (2019) {The Impact of API Evolution on API Consumers and How This Can
  Be Affected by API Producers and Language Designers}. PhD thesis, Delft
  University of Technology

\bibitem[{Sawant et~al.(2016)Sawant, Robbes, and
  Bacchelli}]{Sawant2016Reaction}
Sawant AA, Robbes R, Bacchelli A (2016) On the reaction to deprecation of 25,
  357 clients of 4+1 popular java apis. In: 2016 {IEEE} International
  Conference on Software Maintenance and Evolution, {ICSME} 2016, Raleigh, NC,
  USA, October 2-7, 2016, {IEEE} Computer Society, pp 400--410,
  \doi{10.1109/ICSME.2016.64}

\bibitem[{Wohlin et~al.(2000)Wohlin, Runeson, H\"{o}st, Ohlsson, Regnell, and
  Wessl{\'e}n}]{Wohlin2000Experimentation}
Wohlin C, Runeson P, H\"{o}st M, Ohlsson MC, Regnell B, Wessl{\'e}n A (2000)
  {Experimentation in Software Engineering: An Introduction}. Kluwer Academic
  Publishers, Norwell

\bibitem[{Wu et~al.(2016)Wu, Khomh, Adams, Gu{\'{e}}h{\'{e}}neuc, and
  Antoniol}]{Wu2016Exploratory}
Wu W, Khomh F, Adams B, Gu{\'{e}}h{\'{e}}neuc Y, Antoniol G (2016) An
  exploratory study of api changes and usages based on apache and eclipse
  ecosystems. Empir Softw Eng 21:2366--2412, \doi{10.1007/s10664-015-9411-7}

\bibitem[{Xavier et~al.(2017{\natexlab{a}})Xavier, Brito, Hora, and
  Valente}]{Xavier2017Historical}
Xavier L, Brito A, Hora AC, Valente MT (2017{\natexlab{a}}) Historical and
  impact analysis of {API} breaking changes: {A} large-scale study. In: {IEEE}
  24th International Conference on Software Analysis, Evolution and
  Reengineering, {SANER} 2017, Klagenfurt, Austria, February 20-24, 2017,
  {IEEE} Computer Society, pp 138--147, \doi{10.1109/SANER.2017.7884616}

\bibitem[{Xavier et~al.(2017{\natexlab{b}})Xavier, Hora, and
  Valente}]{Xavier2017Why}
Xavier L, Hora AC, Valente MT (2017{\natexlab{b}}) Why do we break apis? first
  answers from developers. In: {IEEE} 24th International Conference on Software
  Analysis, Evolution and Reengineering, {SANER} 2017, Klagenfurt, Austria,
  February 20-24, 2017, {IEEE} Computer Society, pp 392--396,
  \doi{10.1109/SANER.2017.7884640}

\bibitem[{Xi et~al.(2019)Xi, Shen, Gui, and Zhao}]{Xi2019Migrating}
Xi Y, Shen L, Gui Y, Zhao W (2019) Migrating deprecated {API} to documented
  replacement: Patterns and tool. In: Internetware '19: The 11th Asia-Pacific
  Symposium on Internetware, Fukuoka, Japan, October 28-29, 2019, {ACM}, pp
  15:1--15:10, \doi{10.1145/3361242.3361246}

\bibitem[{Xu et~al.(2019)Xu, Dong, and Meng}]{Xu2019Meditor}
Xu S, Dong Z, Meng N (2019) Meditor: inference and application of {API}
  migration edits. In: Proceedings of the 27th International Conference on
  Program Comprehension, {ICPC} 2019, Montreal, QC, Canada, May 25-31, 2019,
  {IEEE} / {ACM}, pp 335--346, \doi{10.1109/ICPC.2019.00052}

\bibitem[{Zerouali et~al.(2019)Zerouali, Mens, Gonz{\'{a}}lez{-}Barahona,
  Decan, Constantinou, and Robles}]{ZeroualiMGDCR19}
Zerouali A, Mens T, Gonz{\'{a}}lez{-}Barahona JM, Decan A, Constantinou E,
  Robles G (2019) A formal framework for measuring technical lag in component
  repositories - and its application to npm. J Softw Evol Process 31,
  \doi{10.1002/smr.2157}

\end{thebibliography}
